\documentclass[screen, sigconf]{acmart}
\AtBeginDocument{%
  \providecommand\BibTeX{{%
    \normalfont B\kern-0.5em{\scshape i\kern-0.25em b}\kern-0.8em\TeX}}}

\copyrightyear{2025}
\acmYear{2025}
\setcopyright{acmlicensed}\acmConference[CHI '25]{CHI Conference on Human Factors in Computing Systems}{April 26-May 1, 2025}{Yokohama, Japan}
\acmBooktitle{CHI Conference on Human Factors in Computing Systems (CHI '25), April 26-May 1, 2025, Yokohama, Japan}
\acmDOI{10.1145/3706598.3713229}
\acmISBN{979-8-4007-1394-1/25/04}
\usepackage{amsmath}
\usepackage{mathtools}
\usepackage{subcaption}
\usepackage{bbm}

\usepackage{xcolor}
\usepackage{enumitem}

\definecolor{pastelred}{RGB}{255,182,193}
\definecolor{pastelblue}{RGB}{173,216,230}
\definecolor{pastelgreen}{RGB}{144,238,144}

\begin{document}

\title[Contrastive Explanations]{Contrastive Explanations That Anticipate Human Misconceptions Can Improve Human Decision-Making Skills}

\author{Zana Bu\c cinca}
\orcid{0000-0002-2644-6065}
\affiliation{%
  \institution{Harvard University}
  \city{Boston}
  \state{Massachusetts}
  \country{USA}}
\email{zbucinca@seas.harvard.edu}

\author{Siddharth Swaroop}
\orcid{0009-0009-5345-8844}
\affiliation{%
  \institution{Harvard University}
  \city{Boston}
  \state{Massachusetts}
  \country{USA}}
\email{siddharth@g.harvard.edu}

\author{Amanda E. Paluch}
\orcid{0000-0003-4244-9511}
\affiliation{%
  \institution{University of Massachusetts Amherst}
  \city{Amherst}
  \state{Massachusetts}
  \country{USA}}
\email{apaluch@umass.edu}

\author{Finale Doshi-Velez}
\orcid{0000-0003-2886-3898}
\affiliation{%
  \institution{Harvard University}
  \city{Boston}
  \state{Massachusetts}
  \country{USA}}
\email{finale@seas.harvard.edu}

\author{Krzysztof Z. Gajos}
\orcid{0000-0002-1897-9048}
\affiliation{%
  \institution{Harvard University}
  \city{Boston}
  \state{Massachusetts}
  \country{USA}}
\email{kgajos@eecs.harvard.edu}

\renewcommand{\shortauthors}{Bu\c{c}inca, et al.}

\begin{abstract}
People's decision-making abilities often fail to improve or may even erode when they rely on AI for decision-support, even when the AI provides informative explanations. We argue this is partly because people intuitively seek contrastive explanations, which clarify the difference between the AI's decision and their own reasoning, while most AI systems offer ``unilateral'' explanations that justify the AI’s decision but do not account for users' knowledge and thinking. 
To address potential human knowledge gaps, we introduce a framework for generating human-centered contrastive explanations which explain the difference between AI's choice and a predicted, likely human choice about the same task.
Results from a large-scale experiment (N = 628) demonstrate that contrastive explanations significantly enhance users' independent decision-making skills compared to unilateral explanations, without sacrificing decision accuracy. 
As concerns about deskilling in AI-supported tasks grow, our research demonstrates that integrating human reasoning into AI design can promote human skill development.

\end{abstract}

\keywords{AI-assisted decision-making, human-AI interaction, explainable AI, human skills, contrastive explanations}

\ccsdesc[500]{Human-centered computing}
\ccsdesc[500]{Human-centered computing~Empirical studies in interaction design}

\received{12 September 2024}
\received[revised]{10 December 2024}
\received[accepted]{16 January 2025}

\maketitle


\section{Introduction}


\begin{figure*}
\centering
 \includegraphics[scale=0.20]{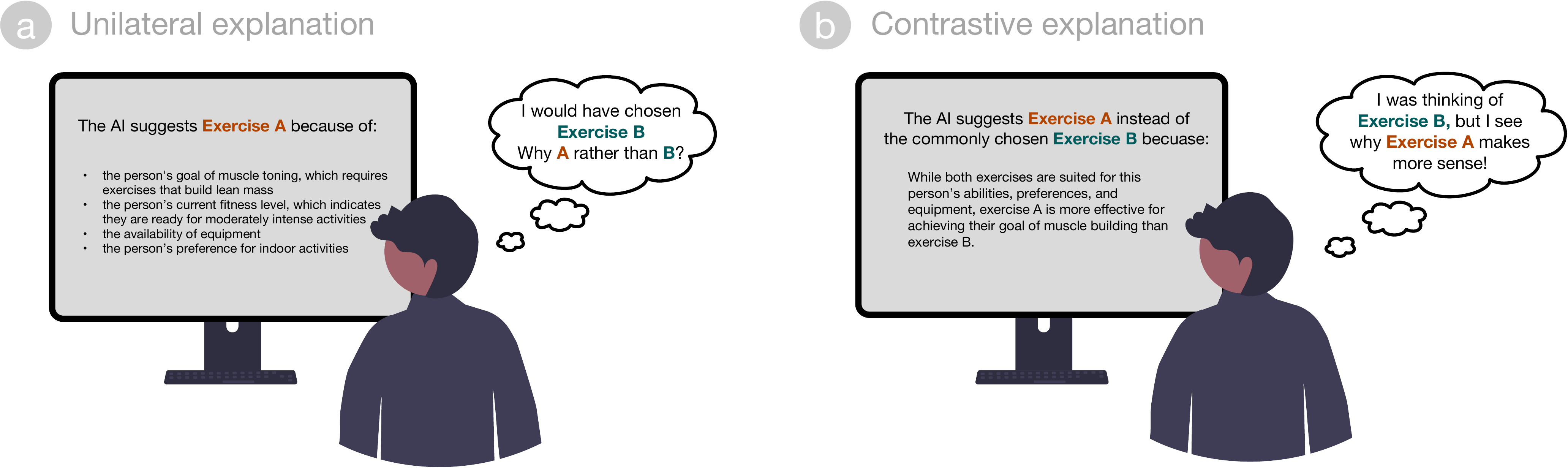}
 \caption{A simplified illustration of (a) unilateral explanations, which list all the features contributing to the AI's decision, and (b) contrastive explanations, which highlight the differences between the AI's choice and a likely human response for an exercise recommendation task.}
 \label{fig:main_figure}
\end{figure*}

Imagine if AI decision-support tools not only improved the quality of our decisions but also enhanced our decision-making skills in the process. Competence, mastery, and skill growth are fundamental drivers of motivation in the workplace and beyond~\cite{deci2012self, deci2017self}. Individuals are inherently driven to refine their abilities in the tasks with which they engage, whether it's making more informed treatment decisions for patients, sharpening writing skills, or mastering a new programming language. The ongoing process of self-improvement not only leads to better outcomes --- decisions, papers, or code --- it also provides intrinsic satisfaction by fulfilling people's fundamental need for competence~\cite{deci2017self}.  As AI systems become more integrated into our decision-making tasks, a critical question arises: How will this assistance affect our skill growth and competence in decision-making? Specifically, as AI systems increasingly offer ready-made ``solutions'', do individuals develop and improve the underlying skills needed to evaluate and generate high-quality decisions independently, or do they risk becoming overly reliant on AI recommendations?

Fueling broader concerns about deskilling~\cite{woodruff2024knowledge, li2024user}, emerging empirical evidence~\cite{rinta2018consequences, gajos2022people, bastani2024generative} suggests that automation and the design of many current AI decision-support systems might not only fail to nurture our skills but could actively degrade them. For instance, \citet{rinta2018consequences} found that people's accounting skill degradation became apparent once a system for fixed assets management was discontinued after years of use. And~\citet{gajos2022people} demonstrated that providing AI-generated recommendations and explanations did not improve people's decision-making abilities, even when those explanations contained facts from which individuals could learn and improve their decision-making abilities.

Some have argued that when people are provided with AI recommendations they may overrely on the recommendation and only superficially process the explanation~\cite{bucinca2021trust, bucinca20:proxy}, thus failing to improve their learning and skills on the task.~\cite{gajos2022people}. Building on this, we posit that even when people attend to the explanation, the reason AI systems fail to improve people's skills, can be partially attributed to the nature of the explanations provided, which often fail to address the specific knowledge gaps that users seek to fill. Typically, these systems offer, what we call, \emph{unilateral} explanations --- justifications that focus on why a particular AI recommendation was made, often by detailing the relevant features~\cite{ribeiro2016should}, highlighting regions~\cite{simonyan2013deep}, or presenting the general reasoning in support of the decision. For example, an AI system might recommend treatment \emph{P} as the best option for a patient because it addresses symptoms X, Y, and Z, without contrasting it to alternative treatments that may address some of those symptoms as well. Yet, research in social science and cognitive psychology has put forth that people naturally seek explanations that are \emph{contrastive} rather than \emph{unilateral}~\cite{miller2019explanation, lipton1990contrastive, hilton1988contemporary}. When people ask why a certain event occurred, or a certain choice was made --- ``Why P?''--- they are often implicitly asking ``Why P (referred to as \textit{fact}) rather than Q (referred to as \textit{foil})?'' --- where foil is a plausible alternative that was considered but not chosen. \citet{jacobs2021designing} found that clinicians would prefer contrastive explanations from AI decision support systems as well, which, for example compare and contrast the AI suggestions with existing standards of care. Such explanations are intuitive and engaging because they focus only on the knowledge gap, addressing the specific points of divergence that are of interest or confusion to the explainee. 

While numerous computational approaches have been introduced for generating contrastive explanations~\cite{alvarez-melis2021from, aguilarpalacios2020cold, van2018contrastive, jacovi2021contrastive}, their focus typically lies on computing the contrast between the fact and the foil rather than on determining a high quality foil. Some approaches consider the foil to be the closest class to the fact in the model~\cite{van2018contrastive}, which we argue results in a model-centric foil that does not necessarily align with human reasoning about the task. Others assume the foil is provided or explicitly inputted by the user~\cite{jacovi2021contrastive}. Such additional step of making initial decisions has been shown to affect both the acceptance of systems and subjective experience in AI-assisted decision-making~\cite{bucinca2021trust, fogliato2022goes}. Thus, predicting a human-centered foil without asking for explicit user input remains a challenge for generating useful contrastive explanations.

Building on the existing insights into AI-assisted decision-making, in this paper, we propose a novel approach to enhance AI-powered decision-support systems by generating human-centric contrastive explanations. Our method leverages a ``mental model'' of humans to provide an explanation in the form of ``Why P rather than Q?'' --- where P represents the AI's recommendation and Q is a plausible alternative response from a human perspective. Unlike unilateral explanations, which justify a recommendation by listing all dimensions that contributed to a decision, our contrastive explanations focus on the distinctions between the AI's suggestion and a likely human response, while highlighting only the dimensions in which the two choices differ. We hypothesize that such contrastive explanations, which highlight knowledge gaps between AI and predicted human responses, will foster greater cognitive engagement and enhance task learning compared to unilateral explanations, while maintaining similar decision accuracy. Additionally, we explore how the quality of the foil (predicted vs. random) and timing (before the person makes a decision vs. after an initial decision) of the contrastive explanation affect their effectiveness, hypothesizing that high-quality foils will maximize learning outcomes and pre-decision timing will maximize acceptance.

To generate contrastive explanations with which to test the hypotheses, we introduce a human-centric framework, which we instantiated for an exercise recommendation decision-making task. Our framework consists of four modules: (1) an AI task model that predicts a response to a decision task (fact), (2) a human model that predicts an average human's response for the same task (foil), (3) a contrast module that identifies the relevant dimensions where the fact and foil differ, and (4) an LLM-powered presentation module that formats these differences into an explanation and adds common sense knowledge (within the constraints provided by the other modules) that bridges the knowledge gap between the AI's recommendation and the human response. To test our hypotheses, we conducted an online between-subjects experiment (N=628) comparing five conditions: no AI, unilateral explanations, contrastive explanations with a predicted foil, contrastive explanations with a random foil, and contrastive explanations provided after an initial decision was made (inputted foil). Our results demonstrated that contrastive explanations with a predicted foil enhanced human skill on the task (i.e., human learning) significantly more than unilateral explanations, without sacrificing decision accuracy. Within contrastive conditions, we found that timing of contrastive explanations affected subjective experience but not objective outcomes. Participants in the contrastive explanations with predicted vs. inputted foil did not differ significantly in terms of decision accuracy or human learning but contrastive explanations with predicted foils resulted in significantly higher subjective perceptions of competence, autonomy, and relatedness to the AI than contrastive explanations with inputted foils. Further, we found that the quality of the foil matters: although we used a single model to predict human responses, participants interacting with contrastive explanations featuring a predicted foil improved their learning more than those with a random foil, though the difference was only marginally significant. This result suggests that personalized foil models, fine-tuned for each individual, may offer additional benefits. 

Our study explored only the short-term effects of explanations on learning. Further research is needed to understand whether these positive effects persist over time.

In summary, this paper makes the following main contributions:
\begin{itemize}
    
    \item We introduced a contrastive explanations framework for generating human-centered contrastive explanations which compare AI's decision choice to a predicted human response for the same task.
    \item Our results demonstrated that such human-centered contrastive explanations significantly enhance decision-making skills without sacrificing decision accuracy compared to unilateral explanations, a default explanation design in AI-powered decision support.
    \item We further presented evidence about which design aspects of contrastive explanations affect objective outcomes and people's intrinsic motivation to engage with the decision task.
    \item Our work is the first to demonstrate that the \emph{content} of explanations significantly impacts the improvement of human skills, opening up new opportunities for developing more effective explanation designs.
    \item Our research suggests that decision support tools that consider the decision-makers' knowledge and mental model of the task can enhance people's understanding and proficiency in the task more effectively than current designs of decision-support which provide AI-centric unilateral explanations. With the growing adoption of AI-powered support across tasks and settings, we believe that our findings may offer a path forward toward AI systems that upskill, augment, and improve human capabilities.
\end{itemize}

\section{Related Work}

\subsection{Contrastive Explanations}

The field of Explainable AI (XAI) has developed a wide range of methods aimed at making AI systems more understandable and useful to people~\cite{gunning2019darpa}. Seminal approaches include feature-based explanations like LIME~\cite{ribeiro2016should} and SHAP~\cite{lundberg2017unified}, which demonstrate how individual features influence an AI decision, as well as saliency maps~\cite{simonyan2013deep}, which highlight image regions that contributed to the outcome. These methods, which we refer to as \emph{unilateral} approaches, focus on explaining why the AI made a specific decision but do so in isolation, without explicitly comparing it to other plausible alternatives.

Meanwhile, \citet{miller2019explanation}'s extensive review of social science research has underscored the significance of contrastive explanations, sparking a new line of inquiry in ML and HCI research. Miller's review highlights that, according to social science literature, explanations people seek and provide are predominantly contrastive~\cite{miller2019explanation, lipton1990contrastive, lewis1986causal}. Rather than simply asking ``Why P?'' to receive a list of features or a sequence of causal events, people often want to know ``Why P instead of Q?'' --- seeking an explanation that clarifies the difference between the actual outcome and an (often implicit) alternative they expected. \citet{lipton1990contrastive} refers to ``P'', the actual event, as the \emph{fact}, and the alternative ``Q'' as the \emph{foil}.

Social science experts emphasize the value of contrastive explanations for two main reasons~\cite{miller2021contrastive}. Firstly, they arise from a person's surprise over an unexpected event, revealing their preconceived expectations --- essentially offering insight into the individual's mental model and highlighting their knowledge gaps~\cite{lewis1986causal, miller2018contrastive}. Secondly, providing and asking for contrastive explanations is less complex and cognitively demanding, making the process more efficient for both the inquirer and the respondent~\cite{lipton1990contrastive, lewis1986causal, ylikoski2007idea}. In AI-assisted decision-making, we further hypothesize that because contrastive explanations highlight (1) the knowledge gap of the inquirer and (2) are shorter, and thus easier to parse, they will result in improved knowledge acquisition from the decision-maker compared to explanations that highlight all the decision factors.

In recent years, machine learning scholars have introduced various computational approaches for generating contrastive explanations, such as pairwise class comparisons~\cite{alvarez-melis2021from, aguilarpalacios2020cold}, tree-based methods~\cite{van2018contrastive, sokol2020one}, or identifying pertinent positives and negatives~\cite{dhurandhar2018explanations}. Unlike in our work in which the foil seeks to convey explainee's thinking and is generated by a separate model, in the existing techniques, the foil is commonly determined as the closest alternative outcome that would alter the model's decision. For example, in tree-based approaches, foils are selected as the closest non-matching class leaf, while in counterfactual reasoning~\citet{hendricks2018generating}, foils are chosen based on their proximity to the input data but belonging to a different class. It is not clear, however, if these methods correctly anticipate how people and the models disagree. We argue that contrastive explanations in which the foil presents the decision-maker's reasoning more accurately reflect the social science understanding of contrastive reasoning, which seek to clarify the gaps in the explainee's reasoning.

One example of contrastive explanations in HCI literature is \citet{zhang2020effect}'s framework for generating contrastive explanations in vocal emotion recognition. Like other machine learning techniques, this framework highlights the differences between two similar instances with different class labels, using high-level, human-interpretable concepts rather than granular features. Other related systems produce counterfactual explanations which compare decision instances or hypothetical input space changes~\cite{jin2020carepre, wang2021gam}, rather than outcome differences. It is important to note that contrastive explanations are often conflated with counterfactual explanations, which explore how minimal input changes could lead to different outcomes, while contrastive explanations clarify differences between two outcomes (e.g., ``Why treatment P rather than Q?''). In multi-class settings, these explanations (contrastive and counterfactual) are distinct, but for binary classifications tasks, the distinction blurs. 

\subsection{AI-Assisted Decision-Making}

\subsubsection{Why optimizing human decision-making skills in AI-assisted decision-making matters?}

A growing concern, especially with the recent developments in generative AI~\cite{woodruff2024knowledge, li2024user}, deskilling refers to the process by which workers lose skills or their proficiency in tasks due to a reduced need to actively engage in those tasks~\cite{braverman1998labor}. This often occurs when technology, such as AI and automation~\cite{susskind2015future}, takes over some or all responsibilities that were previously performed by humans. As individuals become more reliant on these systems to handle complex or repetitive tasks,  they may stop developing or maintaining the expertise required to perform those tasks independently~\cite{arnold1998theory}.  For example, in AI-assisted decision-making, workers might depend on AI to make recommendations or decisions, which can diminish their critical thinking, problem-solving abilities, and overall competence in that domain over time. Indeed, recent empirical evidence shows that the current designs of decision-support tools that provide AI recommendations and explanations do not seem to support people's growth of decision-making skills~\cite{gajos2022people} and evidence from expert-based systems shows that long-term dependence on such systems does lead to deskilling in those very tasks~\cite{rinta2018consequences}.

While powerful, AI systems can be wrong. They make errors due to biases in the data, limitations in the model, or unforeseen circumstances and they even hallucinate. In the short term, when humans have strong decision-making skills, they are better equipped to recognize and override AI mistakes, can critically assess the AI's recommendations, apply domain expertise, and contribute meaningfully to the decision-making process, resulting in more accurate and nuanced outcomes. In the long term, nurturing independent and strong decision-making skills is essential for humans to retain autonomy in decision-making, transfer their expertise to new situations, and adapt to evolving technologies. Such independent decision-making protects both accountability and human agency as AI becomes more integrated into workflows.

Our work adds to the nascent body of research in AI-assisted decision-making, which is concerned with improving human decision-making skills in addition to accuracy of the decisions~\cite{buccinca2024optimizing,gajos2022people}.
While related, AI in education is a distinct field that focuses primarily on utilizing AI to enhance students' learning outcomes and skill development. A notable recent advancement in scalable tutoring involves leveraging large language models (LLMs) as virtual tutors. These models infer student errors and address knowledge gaps by emulating strategies employed by expert tutors~\cite{wang2024bridging}. While some of these approaches may be applicable to decision-making contexts, we believe educational interventions operate under a different set of constraints. For instance, introducing substantial friction in interactions (\emph{e.g.}, as explored in~\cite{kazemitabaar2024exploring}) is often acceptable in education but may not be suitable for real world decision-making scenarios. For a broader overview of related work of AI in education, we direct readers to a recent review~\cite{crompton2024affordances}.

\subsubsection{Eliciting cognitive engagement to calibrate reliance on AI}

Early optimism that AI decision-support tools will inevitably enhance human decision quality~\cite{markus2021role} has dwindled in light of accruing empirical evidence that paints a more complex picture~\cite{bucinca2021trust, wang2021explanations, bansal2021does, schemmer2023appropriate, ghassemi2021false}. Intuitive designs that rely on simple XAI approaches, such as providing AI recommendations alongside (unilateral) explanations, have been shown to lead to overreliance --- where users follow incorrect AI advice --- across diverse tasks, settings, and explanation styles~\cite{bucinca2021trust,gaube2021ai,bansal2021does,jacobs2021how,schemmer2023appropriate}. 
This empirical evidence has prompted extensive research in designing interventions beyond explanation content that encourage appropriate human reliance on AI. Some endeavours to addressing this challenge focus on training or onboarding sessions aimed at helping individuals develop a mental model of the AI~\cite{mozannar2022teaching, mozannar2024effective, pinski2023ai, kawakami2023training, cai2019hello, cai2021onboarding},  providing meta-information about the AI’s uncertainty and limitations~\cite{kim2024m, cabrera2023improving, mozannar2024effective}, helping individuals calibrate their own self-confidence about the task~\cite{he2023knowing, ma2024you}, or enhance user agency by giving them control over input feature selection and algorithmic assistance~\cite{cheng2023overcoming, lai2023selective}. 

By prompting users to reflect on two choices, contrastive explanations fall within one such growing category of interventions designed to compel deeper cognitive engagement with AI support. Scholars have suggested that overreliance on AI often stems from people's superficial engagement with AI recommendations and explanations~\cite{bucinca2021trust, gajos2022people}. In response, various interventions have been developed to enhance cognitive engagement, including cognitive forcing~\cite{bucinca2021trust}, evaluative AI~\cite{miller2023evaluative}, explanations provided without decision recommendations~\cite{gajos2022people}, explanations framed as questions~\cite{danry2023don}, or offering more than one decision suggestion~\cite{lu2024does, chiang2024enhancing}.

While many of these interventions have shown promise in human-AI decision-making quality, they often introduce trade-offs, such as reducing subjective experience~\cite{bucinca2021trust} or requiring more time~\cite{swaroop2024accuracy, cao2023time, swaroop2025personalising}, compared to simply providing AI recommendations with explanations. For instance, cognitive forcing functions~\cite{bucinca2021trust} compel deeper cognitive engagement by requiring people to make decisions before receiving AI support. While these interventions significantly reduce overreliance compared to presenting AI recommendations and explanations upfront, they also lead to significantly lower subjective experience. Empirical evidence suggests that people generally tend to dislike receiving AI support \emph{after} having made a decision~\cite{bucinca2021trust, fogliato2022goes}. Evaluative AI also introduces a paradigm where decision-makers form provisional decisions and then receive AI-generated critiques~\cite{miller2023evaluative}. Albeit, a recent study implemented this concept as an \emph{interactive} interface that allows users to iteratively refine their hypotheses and observe how the evidence aligns or conflicts with their reasoning~\cite{le2024towards}, which appears to preserve the subjective experience compared to receiving static critiques post-hoc. Building on this evidence, we hypothesize that providing static contrastive explanations \emph{after} prompting a person to make an initial decision may similarly have a negative effect on subjective experience, thus hindering the uptake of systems that provide such support in real-life scenarios. However, there is a trade-off here because providing a contrastive explanation after a person has revealed their initial decision has the obvious advantage of revealing the actual foil to the system. This, in turn, can make the explanations more useful for human decision-making and learning compared to settings where the foils are imperfectly predicted.

Offering more than one decision suggestion or source of advice has also been explored as a mechanism to enhance engagement and calibrate reliance on AI support~\cite{bansal2021does, lu2024does, chiang2024enhancing}. For example, \citet{bansal2021does} show that providing top two AI predictions and \citet{lu2024does} show that offering a ``second opinion'' in addition to the main AI support, either from another AI or peers, can reduce overreliance on AI recommendations in certain situations. Contrastive explanations also make \emph{two} options salient to the decision-maker --- the fact and foil --- \textit{along} with reasoning that supports one over the other. It is unclear whether the presence of the explicit comparison in contrastive explanations might dilute the ``second opinion'' effect that previous studies have shown reduces overreliance.

Finally, the studies mentioned above treat cognitive engagement as a mechanism for fostering appropriate reliance on AI, mostly focusing on optimizing human-AI decision accuracy by encouraging deeper thought about AI recommendations. Building on the work of \citet{gajos2022people}, our study instead examines cognitive engagement as a means of enhancing human learning about the task.

\subsubsection{Assisting decision-making with LLM-generated explanations}

The emergence of Large Language Models (LLMs) has sparked interest in their potential to generate explanations that enhance decision-making. In the domain of programming assistants, \citet{yan2024ivie} introduced an LLM-powered system that generates natural language explanations to clarify the functionality of each code suggestion. For data annotation tasks, \citet{wang2024human} leveraged an LLM to predict annotation labels and provide explanations for its choices. In recommendation systems, \citet{silva2024leveraging} used LLMs both as the recommendation engine and as a generator of personalized explanations to improve user experience. In the context of LLM-powered fact-checking, providing contrastive explanations that present evidence on both why a claim might be true and why it might be false has been shown to effectively reduce overreliance on incorrect LLM predictions~\cite{sichengli2024large}.

In contrast to such approaches, which use LLMs for both task-centric predictions (e.g., code suggestions, recommendations) \emph{and} explanation, our work separates these functions. Like \citet{slack2023explaining}, who introduce a chat-based interface to query a predictive machine learning model, we exploit LLMs for their language generation capabilities, and in addition for their common sense reasoning. We rely on a trusted predictive model for generating task-related predictions and explanation dimensions, while the LLM is solely responsible for turning a scaffold produced by the predictive model into natural language, and filling in small common-sense knowledge gaps needed to interpret the model's predictions. This separation preserves the accuracy of predictions and explanations while benefiting from the interpretability and coherence of LLM-generated rationalizations, thereby minimizing the risk of the notorious hallucinations for which LLMs are known~\cite{ji2023towards}.

\subsection{Human Intrinsic Motivation and AI Assistance}

With AI systems redefining workflows and the way tasks are carried out, questions surrounding their effect on people's motivation about the tasks for which they receive assistance are becoming more pressing~\cite{buccinca2024optimizing}. According to the seminal Self-Determination Theory (SDT), individuals feel intrinsically motivated when three psychological needs---competence, autonomy, and relatedness---are met during an activity~\cite{deci2017self}. Competence refers to the need to feel skilled and effective in the activity, autonomy reflects the need to have control over how the activity is carried out, and relatedness involves the need to feel connected to others and to experience a sense of belonging while engaging in the activity. These three needs are fundamental for fostering intrinsic motivation, which leads to greater engagement, performance, and overall satisfaction with the task~\cite{deci2017self}. The introduction of AI assistance into decision-making processes can affect these psychological needs in multiple ways. For example, while AI might enhance short-term feelings of competence by providing support in the moment of decision-making~\cite{fisher2021harder}, it may simultaneously undermine long-term mastery, as current designs do not always facilitate skill development~\cite{gajos2022people}. Similarly, AI can diminish a user's sense of autonomy if they feel overly dependent on the system, reducing their ownership of task outcomes. 

We hypothesize that both the outcomes of the interaction and the design of the AI system influence perceptions of competence and autonomy. On the outcome side, we expect that AI systems that actively support skill development will enhance feelings of competence. In terms of design, approaches where the AI critiques each decision after it is made (\emph{e.g.}, contrastive after) may undermine users' feelings of competence and autonomy, as they could perceive the AI more as a micromanager than a supportive tool, constantly pointing out flaws and dictating its preferred way of doing things. Additionally, even when AI assistance is provided before a decision, designs that emphasize only one option (\emph{e.g.}, unilateral condition) may reduce the sense of autonomy by shifting the decision-maker into a more passive role. In contrast, designs that present multiple options promote active engagement, allowing the decision-maker to weigh different possibilities before making a choice (\emph{e.g.}, contrastive before conditions).

In SDT, relatedness refers to the connection an individual feels toward colleagues or collaborators, typically measured through questions about trust, similarity in reasoning, and willingness to engage in future interactions. We adapt these constructs to assess relatedness to AI, hypothesizing that designs fostering competence and autonomy will similarly enhance relatedness to AI systems.

\section{The Contrastive Explanation framework 
\& Hypotheses}
\label{sec:contrastive_framework}

\label{section:overview}
\begin{figure*}
     \centering
     \includegraphics[width=\textwidth]{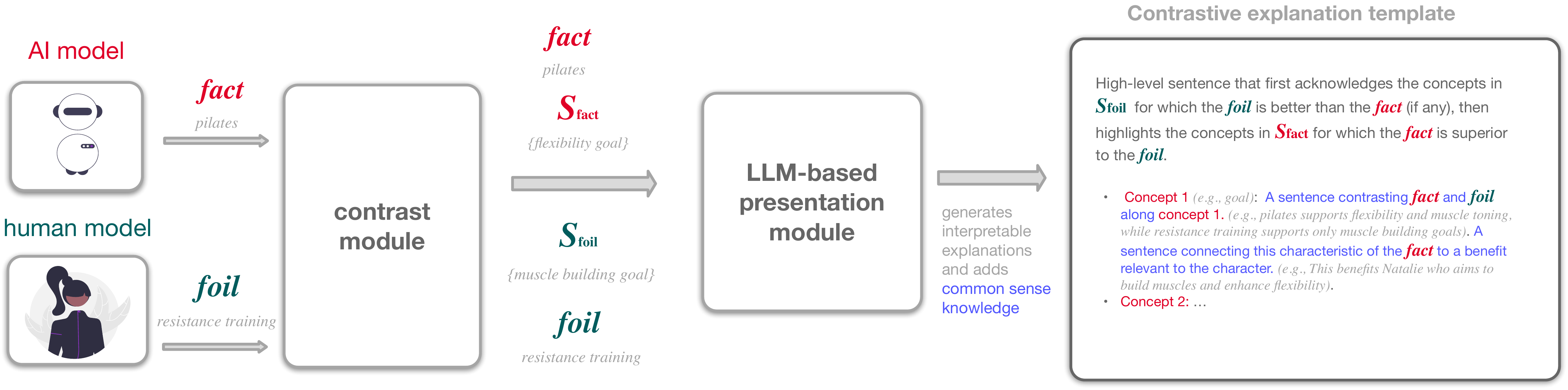}
     \caption{The Contrastive Explanation Framework: The AI task model predicts the AI’s response for a given decision task (fact), while the human model predicts the user’s response for the same task (foil). The contrastive module then analyzes the differences between the AI's and the human's responses, generating task dimensions where the fact is superior to the foil ($S_\textrm{fact}$) and, if any, where the foil is superior to the fact ($S_\textrm{foil}$). Finally, the presentation module, powered by a large language model (LLM), formats the information into an interpretable explanation, filling in small common-sense knowledge gaps within the constraints of the predictive models. The example generation outlined in the figure is relates to the character vignette in Figure~\ref{fig:exercise-task}.}.
     \label{fig:framework}
\end{figure*}

Imagine a clinician reviewing an AI-powered decision-support system's recommendation for a patient's treatment plan. The AI suggests Medication A, but the clinician had Medication B in mind based on their experience with this condition. Existing AI systems would simply explain why Medication A is recommended. However, this leaves the clinician wondering why Medication B, which they deemed suitable, is not the better choice. A contrastive explanation may elucidate this knowledge gap as follows: \textit{``While Medication B is a common and a viable choice for most patients because of its short treatment duration, Medication A is recommended due to its lower risk of drug interactions with this patient's current medications.''}

We propose the Contrastive Explanation Framework to address the limitations of current AI-powered decision-support systems by providing contrastive explanations that acknowledge human's alternative considerations when suggesting a decision. This framework is composed of four main components: (1) an AI task model, (2) a model of how humans are likely to reason about this task, (3) a contrastive module, and (4) a presentation module. The AI task model is the standard AI system that predicts the AI’s response for a given decision task (fact), while the human model predicts an average user's response --- a plausible alternative (foil) --- for the same task based on a model trained on previous human decisions. Based on AI model (e.g., weights), the contrastive module then analyzes the differences between the AI’s and the likely human’s responses, generating task concepts in which the fact is superior to the foil (e.g., lower risk of drug interaction) and task concepts, if any, in which the foil is superior to the fact (e.g., shorter treatment duration). Finally, the presentation module, powered by a large language model, formats these dimensions into prose and fills in the common sense knowledge that focuses on the knowledge gap that may lead someone to pick foil as opposed to fact.

To evaluate the effectiveness of contrastive explanations in improving human learning and accuracy in AI-assisted decision-making, we instantiated this framework with an exercise recommendation task, and conducted an experiment in which people were asked to complete a sequence of decisions and were randomized in one of the 5 different conditions: 

\begin{itemize}
\item\textbf{No AI (Baseline).} Participants in the No AI condition completed the study without any AI support.
\item\textbf{Unilateral} In this condition, participants interacted with the typical AI recommendation and explanation paradigm. The AI suggested a choice and provided reasoning to justify why that choice was the best one. The explanation was unilateral, emphasizing all the concepts and evidence supporting the AI's suggestion.

\item\textbf{Contrastive predicted (with predicted foil).} The \textit{contrastive predicted} condition was designed to provide participants with a contrastive explanation that compares the AI recommendation (fact) with the alternative (foil) that a human may likely consider, as predicted by the human model. In the interface, we presented the foil as a choice that ``many people'' would likely make in a similar situation. The explanation highlighted only the concept(s) in which the two choices differ, emphasizing why the AI's recommendation is superior to the foil.

\item\textbf{Contrastive random (with random foil).} Presentation-wise, the \textit{contrastive random} condition was identical to the contrastive condition. However, in this case, the foil was selected randomly from the six possible choices rather than being predicted by the human model.

\item\textbf{Contrastive after (with inputted foil).} In the \textit{contrastive after} condition, participants first made their own decision before receiving the AI's recommendation and the contrastive explanation, in which participant's decision was used as the foil. In situations when the inputted foil was the same as the AI suggestion, participants were presented with a unilateral explanation supporting their choice.
\end{itemize}

\subsection{Hypotheses and Research Questions}

In our hypotheses, we sometimes refer jointly to \emph{contrastive predicted} and \emph{contrastive after} conditions, in which the foil is not random, as \emph{contrastive with a sensible foil}.

Our main hypotheses are that contrastive explanations with a sensible foil will improve participants' decision-making skills~\footnote{In this paper, we use the terms ``improving human learning'' and ``improving decision-making skills'' interchangeably.} more effectively and result in accuracy that is equal to or better than unilateral explanations. Furthermore, within the contrastive conditions, we hypothesize that \emph{contrastive explanations with a predicted foil} will result in greater human learning than those with a random foil, and offer a superior subjective experience compared to contrastive explanations with an inputted foil.

We categorize these main and other hypotheses and research questions by interaction outcomes --- human learning, accuracy, and subjective experience --- and elaborate them below. To enhance readability, we abbreviate learning-focused hypotheses and research questions as H-L and RQ-L and accuracy-focused ones as H-A and RQ-A, respectively. For hypotheses related to subjective measures, we use the -S suffix (e.g., H-S1).

\subsubsection{Human Learning}


\begin{itemize} [leftmargin=*]
\item[]\textbf{H-L1:} Contrastive explanations with sensible foil --- predicted \textbf{(H-L1a}) or inputted \textbf{(H-L1b)} --- will lead to more learning than providing people with no AI support.

\item[]\textbf{H-L2:} Contrastive explanations with sensible foil --- predicted \textbf{(H-L2a}) or inputted \textbf{(H-L2b)} --- will lead to more learning than unilateral explanations.

\item[]\textbf{H-L3:} Contrastive explanations with predicted foil will lead to more learning than contrastive explanations with a random foil.

\item[]\textbf{RQ-L1:} Will contrastive explanations with predicted foil (provided at the decision-making time) lead to different learning than contrastive explanations after the decision is made (contrastive after)?
\end{itemize}

\subsubsection{Accuracy \& Overreliance}

\begin{itemize}[leftmargin=*]

\item[]\textbf{H-A1:} Contrastive explanations with sensible foil --- predicted \textbf{(H-A1a)} or inputted \textbf{(H-A1b)} --- will lead to equal or better decision accuracy compared to unilateral explanations.

\item[]\textbf{RQ-A1:} Will contrastive explanations --- predicted or random --- which present two choices, reduce overreliance on AI, compared to unilateral explanations? 

\end{itemize}

\subsubsection{Subjective Experience}

\begin{itemize}[leftmargin=*]
\item[]\textbf{H-S1:} Contrastive explanations with predicted foil will lead to higher perceived competence, autonomy, and relatedness to AI than unilateral explanations. 

\item[]\textbf{H-S2:} Contrastive explanations with predicted foil will lead to higher perceived competence, autonomy, and relatedness to AI than contrastive explanations with inputted foil.
\end{itemize}

In the following sections, we describe an exercise recommendation task and the instantiation and implementation of the contrastive explanations framework for the exercise recommendation task.

\section{Exercise Recommendation Task Design}
\label{sec:exercise_task_design}

\begin{figure*}
     \centering
     \begin{subfigure}{0.6\textwidth}
         \centering
         \includegraphics[width=\textwidth]{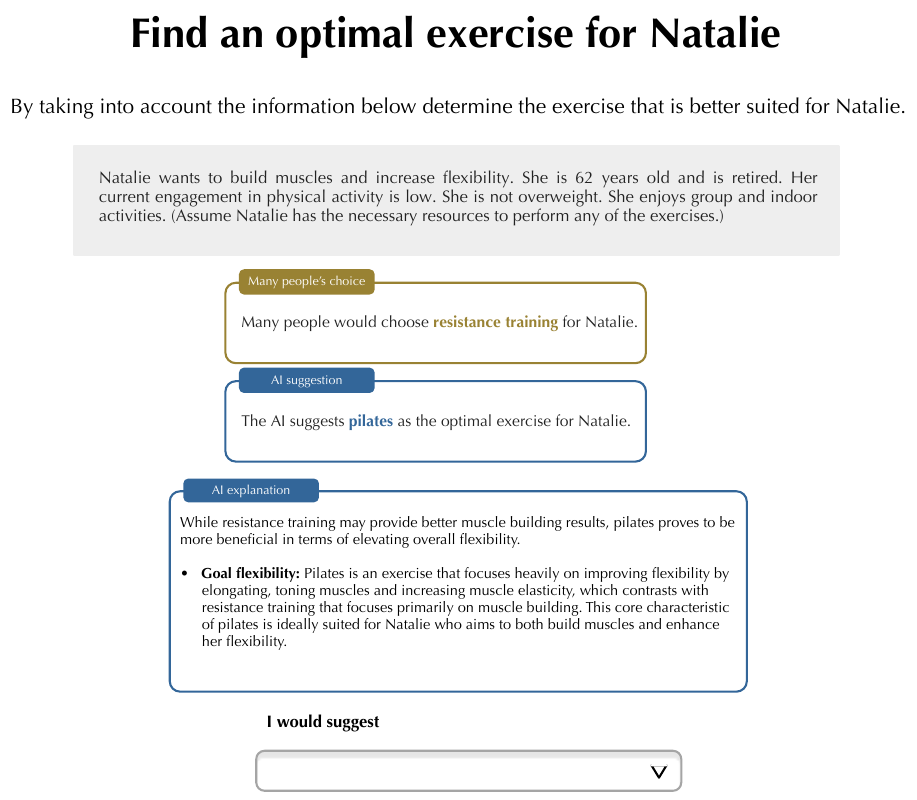}
         \caption{Sample task with contrastive explanation}
         \label{fig:contrastive}
     \end{subfigure}
     \hspace{15px}
      \begin{subfigure}{0.3\textwidth}
       
     \begin{subfigure}{\textwidth}
         \centering
         \includegraphics[width=\textwidth]{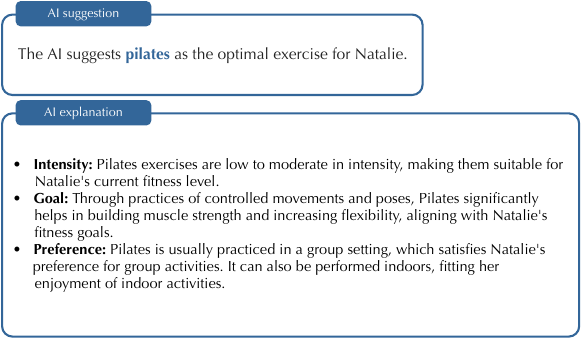}
         \caption{Unilateral explanation}
         \label{fig:single-sided}
     \end{subfigure}

        \vspace{15px}
      \begin{subfigure}{\textwidth}
         \centering
         \includegraphics[width=\textwidth]{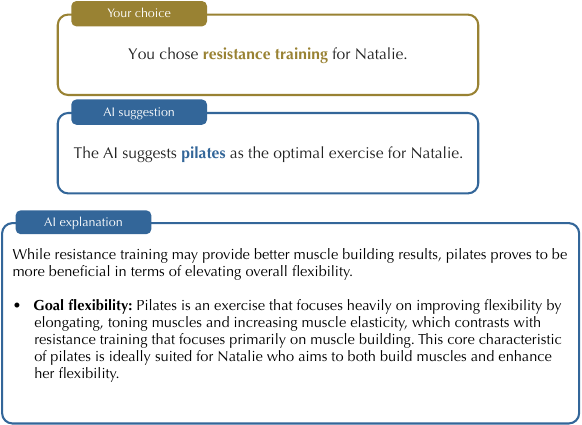}
         \caption{Contrastive explanation after}
         \label{fig:contrastive-after}
     \end{subfigure}
      \end{subfigure}

     \caption{Illustration of the exercise recommendation decision-making task featuring different explanation designs. \ref{fig:contrastive} shows a sample of the task with \textit{contrastive} explanation, whereas ~\ref{fig:single-sided} and \ref{fig:contrastive-after} depict only the explanations for the respective conditions. In the \textit{contrastive random} condition, the presentation was identical to the contrastive condition, but with the alternative (foil) selected randomly. In the \textit{no-AI} condition (not illustrated), participants made decisions without any AI assistance.} 
     \label{fig:exercise-task}
\end{figure*}

To create a decision-making task accessible to laypeople on crowd-sourcing platforms while presenting cognitive challenges similar to high-stakes decisions (e.g., treatment selection), we collaborated with a kinesiology expert, a co-author of this paper. We designed scenarios for an exercise recommendation task, as shown in Figure~\ref{fig:exercise-task}. Participants are tasked to choose the best exercise from a list of options based on a fictional character's description, goals, and preferences. This task is designed to be easy to understand yet complex enough to mimic clinical treatment decisions. Clinicians consider many (sometimes competing) factors when selecting treatments, such as patient condition, treatment preferences, side-effect tolerance, and constraints. Similarly, selecting the right exercise involves weighing the individual's goals, preferences, and capabilities, requiring analogous cognitive steps.

\subsection{Generating the fictitious characters}

We generated vignettes of fictitious people by randomly sampling their demographics from probabilities obtained from the US Census\footnote{data.census.gov}, Centers for Disease Control and Prevention\footnote{data.cdc.gov}, and the US Bureau of Labor statistics\footnote{bls.gov} (name, age, gender, BMI, physical activity level, occupation). According to the sampled fictitious character, we manipulated or randomly sampled the following factors which were deemed important for exercise prescription by the expert: (1) their fitness level and maximal intensity (based on demographics and health status), (2) their exercise goal (\em e.g.\em, building muscles, weight loss, flexibility), and (3) their exercise preference (\em e.g.\em, indoor/outdoor, group/individual). We implemented these steps as fictitious character generation process that allowed us to generate different characters.

\subsection{Curating the exercises}

To build an exercise repository for recommending activities to fictional individuals, we curated a list of 59 leisure activities from a comprehensive compendium, which included various physical activities, from sports to everyday tasks like housework and occupational activities~\cite{ainsworth20112011}. In the compendium, each activity was labeled with its MET (metabolic equivalent), which denotes the energy requirement for basal homeostasis (1 MET is roughly the energy required to sleep or watch TV). Moderate activities require between 3 and less than 6 METs, while vigorous activities require 6 or more METs. We also labeled the exercises based on (i) their goals (cardio, muscle building, flexibility), (ii) whether they are typically performed indoors or outdoors, and (iii) whether they are typically performed individually or in a group. From this list, we selected seven representative exercises for the dropdown menu: aerobics, bicycling, boxing, jog/walk combination, pilates, resistance training, and swimming. See Appendix~\ref{appx:selecting-drop-down-exercises} for a detailed description of the selection process.

\subsection{Representing characters and exercises}

To prescribe exercises to characters, we first represented exercises and characters in a joint representation space. Guided by the domain expert, we constructed a relatively simple representation space consisting of three broad concepts: (1) intensity, (2) goal, and (3) preference.

Each exercise and generated character was encoded onto these three broad concepts as described below.

\textbf{Intensity.} 
For exercises, intensity captures the level of exertion or effort the exercise requires, measured in METs. One MET is defined as the oxygen consumption of 3.5 milliliters of oxygen per kilogram of body weight per minute (3.5 ml/kg/min), which is roughly the rate of oxygen consumption at rest.  

For characters, intensity captures the level of exertion or effort a character can sustain during physical activity (i.e., their cardiorespiratory fitness). It is quantified by the reserve oxygen uptake ($VO_{2_{R}}$), which represents the additional oxygen consumption capacity a person has beyond their resting state. This reserve is determined by subtracting the resting oxygen uptake (3.5 ml/kg/min or 1 MET) from the maximal oxygen uptake ($VO_{2_{max}}$), which is the highest rate at which the body can use oxygen during intense physical activity. Maximal oxygen uptake is assessed in clinical settings using a treadmill test, but various equations have been proposed as useful proxies~\cite{peterman2020accuracy}. Following~\citet{jang2012estimation}, we calculated the cardiorespiratory fitness of a character as a function of age, sex, BMI, and current physical activity level~\footnote{$VO_{2_{max}} =48.392 - 0.088 (age) + 12.335 (sex; men=1, women=0) - 0.386 (BMI) + 0.693 (PA)$}(rating of physical activity on a 7-point scale~\cite{jackson1990prediction}).

\textbf{Goal.} 
For exercises, goal captures the type of benefit the exercise has on the body, consisting of three dimensions: cardiovascular improvement, muscle building, or flexibility.
For characters, goal reflects what the character aims to achieve through exercise, in terms of the same three dimensions: improving cardiovascular health, building muscle, or enhancing flexibility. Note that additional domain knowledge (e.g., cardio is beneficial for weight loss) is necessary to convert some of the higher level character's exercise goals (e.g., losing weight) to the representation space.

\textbf{Preference.}
For exercises, preference indicates whether the exercise is typically performed indoors or outdoors, and whether it is usually done individually or in a group. For characters, preference captures the character's preferred exercise environment (indoor/outdoor) and social setting (individual/group).

\subsection{Designing the objective function}
\label{subsec:designing-objective-function}

Having constructed joint representations for characters and exercises, we now formalize our setting and explain the objective function we designed for recommending exercises to characters.

Let a fictitious character representation be $\mathbf{x} \in \mathbb{R}^{D}$ and an exercise representation be $\mathbf{y} \in \mathbb{R}^{D}$, where $D=6$ and both representations are structured with dimensions representing intensity, goals, and preferences: $$\mathbf{x} = [x_{\textrm{MET}}, x_{\textrm{cardio}}, x_{\textrm{muscle}}, x_{\textrm{flexibility}}, x_{\textrm{environment}}, x_{\textrm{social setting}}]^T,$$ with $\mathbf{y}$ following a similar structure. Our goal was to create a function that scores the ``goodness'' of an exercise for the given character. We designed a linear objective function: 
\begin{equation}
    f(\mathbf{g}(\mathbf{x, y}), \mathbf{w}) = \mathbf{w}^T\mathbf{g}(\mathbf{x,y}),
\label{eq:objective-function}
\end{equation}
where $\mathbf{g}(\mathbf{x, y})$ is a piece-wise vector-valued function (devised with the expert) that returns a joint representation (vector) of the person and the exercise for each dimension. $\mathbf{g}(\mathbf{x, y})  \in \mathbb{R}^{D+1}$ concerning the following aspects: intensity, goal, and preference.

\begin{equation}
\mathbf{g}(\mathbf{x}, \mathbf{y}) = 
\begin{bmatrix}
\textcolor{black}{\min(0, x_1 - y_1)} \\[0pt]
\textcolor{gray}{\text{\textit{\scriptsize\textbf{Intensity}: Penalize exercises exceeding character's capabilities.}}} \\[6pt]
\textcolor{black}{\min(0, y_1 - x_1)} \\[0pt]
\textcolor{gray}{\text{\textit{\scriptsize\textbf{Intensity}: Penalize exercises underutilizing character's capabilities.}}} \\[6pt]
\textcolor{black}{[\mathbbm{1}[x_c > 0]((y_c - x_c) + \mathbbm{1}[y_c == x_c])]_{c\in\{2, 3, 4\}}} \\[0pt]
\textcolor{gray}{\text{\textit{\scriptsize\textbf{Goal}: Match each stated subgoal (cardio, muscle building, flexibility).}}} \\[6pt]
\textcolor{black}{\mathbbm{1}[y_c == x_c]_{c\in\{5, 6\}}}\\[0pt]
\textcolor{gray}{\text{\textit{\scriptsize\textbf{Preference}: Match each preference (environment, social setting).}}} \\[6pt]
\end{bmatrix}
\end{equation}

In the equation above, the subscripts refer to dimensions of $\mathbf{x}$ and $\mathbf{y}$, and $\mathbbm{1}$ denotes the indicator function, which takes the value 1 if the condition inside the brackets is true and 0 otherwise.

\subsection{Learning the expert weights}
\label{subsec:learning-expert-weights}

The parameterized objective function (equation~\ref{eq:objective-function}) enables learning weights $\mathbf{w}$ from different sources of labels. We aim to learn $f_{\text{expert}}$ with weights $\mathbf{w}_{\text{e}}$ based on expert labels, and $f_{\text{human}}$ with weights $\mathbf{w}_{\text{h}}$ based on crowdworker labels. The expert model $f_{\text{expert}}$, takes a description of a fictitious character and exercise, and outputs a real-valued score indicating how well the exercise matches the goals, abilities and preferences of the fictitious character. We trained and validated this model on expert labels of optimal exercises for a series of characters. Similarly, the human model, which captures how humans reason on average (described in ~\ref{subsec:human-model}), was trained on crowdworkers' (i.e., laypeople's) labels.

Generating a series of diverse fictitious characters, we asked the kinesiology expert to select top exercises for them from a list of top 15 exercises (out of 59) selected with a ``dummy'' scoring function which equally weighted each dimension. For every character, the expert provided a best set of exercises  $\mathcal{S}_{1}$ typically consisting of two or three similar exercises (e.g., pilates, yoga), and a second best set of exercises  $\mathcal{S}_{2}$ that would still be a reasonable choice but not as good as the first set.
With 15 exercises in the list, these labels provided multiple pairwise comparisons between the individual exercises. For every exercise $\mathbf{y}^i \in \mathcal{S}_{1}$, then $\mathbf{y}^i$ is a better choice than $\mathbf{y}^j$ for every other exercise $\mathbf{y}^j \notin \mathcal{S}_{1}$. Similarly, for every exercise $\textbf{y}^i \in \mathcal{S}_{2}$, then $\mathbf{y}^i$ is a better choice than $\mathbf{y}^j$ for every $\mathbf{y}^j \notin \{\mathcal{S}_{1} \bigcup \mathcal{S}_{2}\}$.

With a dataset of rankings, our goal was to learn the expert weights $\mathbf{w}_e$ from equation~\ref{eq:objective-function}. Ranking problems, particularly with linear ranking functions, can be transformed into classification problems by considering pairwise differences between elements~\cite{herbrich2000large}. This approach involves transforming the ranking task of a set of items (e.g., exercises) into several binary classification tasks. For each item pair, a difference vector of their features ($\mathbf{u}^i - \mathbf{u}^j$) is generated and the label corresponds to their relative order (e.g., the label $v = 1$ if item $i$ is a better choice than item $j$ and $-1$ otherwise). In our setting, the items correspond to exercises. A binary classifier is then trained on these labeled pairs to predict which of the given two items should be ranked higher. When using a linear binary classifier $v = sign(\mathbf{w}^T (\mathbf{u}^i - \mathbf{u}^j) + b)$, the coefficients of the model represent the weights of the feature differences, thereby indicating the importance of each feature dimension in determining the ranking.

Let exercise $\mathbf{y}^*$ be a better choice than exercise $\mathbf{y}^i$ for a character $\mathbf{x}$. In our setting this looks as follows:
    
$$[\mathbf{g}(\mathbf{x, y}^*) - \mathbf{g}(\mathbf{x, y}^i), 1] \text{ or } [\mathbf{g}(\mathbf{x, y}^i) - \mathbf{g}(\mathbf{x, y}^*), -1],$$ 
where the first element in the square brackets is the input to our classifier model, and the second element is the label. 

To avoid biasing the classifier, we randomly assign a pair to either have a positive (1) or a negative (-1) label (i.e., ``$\mathbf{y}^*$ is better than $\mathbf{y}^i$'' or ``$\mathbf{y}^i$ is worse than $\mathbf{y}^*$'').
We fit an SVM classifier with a linear kernel to these tuples of data with expert labels, thereby recovering the coefficients as the expert weights $\mathbf{w}_e$ for the scoring function: $f(\mathbf{g}(\mathbf{x,y}), \mathbf{w}_e) = \mathbf{w}_e^T \mathbf{g}(\mathbf{x,y})$. (For implementation details and the evaluation of the expert model see Appendix~\ref{appx:evaluating-exper-model}.) We followed a similar approach to learn the human model weights from crowd-sourced data, as described in Section~\ref{subsec:human-model}.

\section{Applying the Contrastive Explanation Framework to the Exercise Task}

Our goal is to generate contrastive explanations (using the framework in Section \ref{sec:contrastive_framework} for the exercise recommendation task explained in Section \ref{sec:exercise_task_design}). 
In this section, we describe this process, and we end up with contrastive explanations like the ones shown in Figure \ref{fig:exercise-task}. 
To do so, we use a simulated AI model (we control the accuracy of this model), generate foils using the human model weights, generate contrast concepts using our representation \(\mathbf{g}(\mathbf{x}, \mathbf{y})\), and generate the explanations using an LLM. 

\subsection{Simulated AI model: Generating the \textit{fact}}
\label{subsec:ai-model}

The AI model in our framework represents the common way in which models are trained for specific tasks (e.g., disease diagnosis) by exposing them to vast amounts of data, which allows them to identify patterns and make decisions based on learned statistical relationships. However, because these models operate solely within the confines of the data they have encountered, they achieve high performance in familiar decision instances, but they also make mistakes when encountering novel or poorly-represented scenarios. 

To emulate real-world situations, we designed a \emph{simulated} AI model such that it performs better than the average human but that it also occasionally makes mistakes. We chose to simulate the AI model because we wanted to have control over the types of mistakes the AI makes. Our formative studies indicated that unassisted people achieve on average 30\% accuracy on selecting the top exercise out of 7 choices, and we designed the AI model to have an accuracy 71.4\%. We used the expert model weights to decide the top exercise recommendation when the AI made a correct decision. For a given decision instance (i.e., character), the top AI suggestion is the exercise with the highest score under the expert model weights $\mathbf{y}_{\textrm{fact}} = \text{argmax}_{\mathbf{y}^i}(f(\mathbf{g}(\mathbf{x},\mathbf{y}^i), \mathbf{w}_{e})$. In the contrastive explanation framework, we refer to the AI generated exercise as the \textit{fact} (even when it is a wrong suggestion). 

To make the AI err, we chose to select a reasonable alternative from among the exercises rather than a random one, as the latter would make AI errors too obvious to participants. Therefore, the AI suggestion in such instances was the foil --- the top exercise selected by the human model (as described in Section~\ref{subsec:human-model}: this is always different to the expert model's top exercise). When the AI errs, the new foil becomes the second-best choice from the human model. As a result, in those instances, neither the fact nor the foil corresponds to the correct answer.

\subsection{Human model: Generating the \textit{foil}}

\label{subsec:human-model}
As motivated in previous sections, we believe that contrastive explanations are most effective when the foil represents a likely human answer. 
For instance, in contexts with established guidelines, such as medical decision-making, the foil could be the guideline-recommended action~\cite{jacobs2021designing}. In situations without established guidelines, the foil can be inferred from prior human decisions. 
In our implementation of the contrastive explanation framework for the exercise recommendation task, we chose to implement the \textit{foil} as the likely human response to a given question. Specifically, we build a human model that predicts the exercise laypeople would select for previously unseen fictitious characters by training on unassisted human responses. We implemented a generic model to represent human decision-making, which was sufficient for our simple task. However, depending on the context, personalized models that adapt and update as they learn more about individual users could be more appropriate.

We generated a series of fictitious characters and ran an online study on Prolific to collect responses from crowd workers who served as non-domain experts. See Appendix~\ref{appx:data-collection} for details of the data collection study and the evaluation of the human model. To learn the human model weights, we followed the same procedure as we did for the expert model weights, and as described in section \ref{subsec:learning-expert-weights}. 

Given a character and two exercises, our learned linear SVM classifier predicted which exercise is more likely to be selected by the human non-expert. 
The coefficients of the classifier with which this decision was achieved yielded the human model weights for each concept (i.e., goal, intensity, preference). Therefore, we constructed a scoring function based on human model weights as well: $f(\mathbf{g}(\mathbf{x, y}), \mathbf{w}_h) = \mathbf{w}_h^T\mathbf{g}(\mathbf{x,y})$.

In our implementation, we selected the foil as the exercise with the highest score under the human weights that was not the same as the expert choice: $\mathbf{y}_{\textrm{foil}} = \text{argmax}_{\mathbf{y}^i}(f(\mathbf{g}(\mathbf{x},\mathbf{y}^i), \mathbf{w}_h)$, where $\mathbf{y}^{i} \neq \mathbf{y}_{\textrm{fact}}$. This approach selects the most likely \emph{incorrect} human answer. When the simulated AI was to provide a wrong suggestion (i.e., the \textit{fact} was suboptimal), the output of this human model was presented as the \textit{fact}, and the new \textit{foil} was the second most likely incorrect human model answer: this is still an incorrect choice, but less likely to be selected by people than the first one.

\subsection{Contrast Module: Generating the contrast concepts}

The goal of the contrast module is to generate the dimensions or features in which the fact and the foil differ. Specifically, what aspects render the fact superior to the foil, and in what aspects (if in any) is the foil superior to the fact.

In our setting, these dimensions indicate the three main concepts of the task: \textit{intensity}, \textit{goal}, and \textit{preference}.
To generate these dimensions we employed the following approach. Let \(\mathbf{y}_{\textrm{fact}}\) and \(\mathbf{y}_{\textrm{foil}}\) be the two exercises generated by the AI and the human model for character \(\mathbf{x}\), respectively. Our goal is to identify the dimensions in which these two exercises differ based on the expert model's weights. For each exercise, we computed the element-wise product of the expert model weights with the joint character-exercise representation \(\mathbf{g}(\mathbf{x}, \mathbf{y})\), resulting in the weighted vectors \(\mathbf{w}_{e} \circ \mathbf{g}(\mathbf{x}, \mathbf{y}_{\textrm{fact}})\) and \(\mathbf{w}_{e} \circ \mathbf{g}(\mathbf{x}, \mathbf{y}_{\textrm{foil}})\) for the AI-generated exercise and the human model-generated exercise, respectively.

Next, we calculated the difference between these two weighted vectors to determine the dimensions along which the exercises differ according to the expert model's weighting scheme. This difference vector, \(\mathbf{\Delta g}_{AI}\), is given by:

\begin{equation}
\mathbf{\Delta g}_{AI} = \mathbf{w}_{e} \circ \mathbf{g}(\mathbf{x}, \mathbf{y}_{\textrm{fact}}) - \mathbf{w}_{e} \circ \mathbf{g}(\mathbf{x}, \mathbf{y}_{\textrm{foil}})
\end{equation}

Non-zero dimensions of $\mathbf{\Delta g}_{AI}$ indicate where the two exercises differ. A positive value indicates that the fact is superior to the foil in that dimension, while a negative value indicates that the foil is superior to the fact.
Therefore, the contrastive module generates two sets of dimensions, dimensions for which the fact is superior to the foil: \(\mathcal{S}_{\textrm{fact}} = \{c \mid \mathbf{\Delta g}_{AI}[c] > 0\}\) and those for which the foil is superior to the fact \(\mathcal{S}_{\textrm{foil}} = \{c \mid \mathbf{\Delta g}_{AI}[c] < 0\}\), where $c$ denotes the dimension. Because the foil may not be superior to the fact in any dimension,  $\mathcal{S}_{\textrm{foil}}$ can be an empty set. However, by definition $\mathcal{S}_{\textrm{fact}} \neq \emptyset$.

\subsection{Presentation Module: Generating interpretable explanations}

Once the \textit{fact}, \textit{foil}, and the dimensions where they differ are generated, the presentation module's purpose is to convert this information into a format that is easily understood by humans. We chose to implement an LLM-powered presentation module which is guided by our trusted predictive model, allowing little room for hallucinations. Given $\mathbf{y}_{\textrm{fact}}$, $\mathbf{y}_{\textrm{foil}}$, and the sets for which each are superior ($\mathcal{S}_{\textrm{fact}}$, $\mathcal{S}_{\textrm{foil}}$), the LLM-powered presentation module adds common sense knowledge and turns the explanations into prose. 

Specifically, the LLM adds knowledge to create the mapping from the the representation space (i.e., concepts) in which the predictive model operates to the input (i.e., vignette) and output spaces (i.e., exercises).
For example, let $\mathbf{x}$ be a fictitious character whose goal is to lose weight. Let $\mathbf{y}_{\text{fact}}$ correspond to the representation of activity \textit{running} and $\mathbf{y}_{\text{foil}}$ correspond to the representation of activity \textit{pilates}. Further, let $\mathcal{S}_{\text{fact}}$ include \{\textit{goal\_cardio}\}. In other words, \textit{running} is superior to \textit{pilates} because it supports \textit{cardio goals}. The remaining domain knowledge required to fully understand the explanations are the following: `cardio benefits weight loss', `running is a cardio exercise' and `pilates is not a cardio exercise'. 

Therefore, highlighting cardio as a differing dimension may not be enough without explaining those domain facts. 
The LLM is prompted to fill in these knowledge gaps, given the information `\emph{running} is superior to \emph{pilates} in supporting \emph{cardio} goals'. 
Note that there is little room for the LLM to hallucinate facts, because we are constraining the generation process with the fact, foil, and concepts ($\mathcal{S}_{\textrm{fact}}$, $\mathcal{S}_{\textrm{foil}}$) that are generated by the predictive models. 

The LLM was always shown the character's vignette, and told the representation space dimensions that we identified as important (from Section~\ref{subsec:designing-objective-function}).

As shown in Fig~\ref{fig:framework}, for contrastive explanations, the LLM was given $\mathbf{y}_{\textrm{fact}}$, $\mathbf{y}_{\textrm{foil}}$, $\mathcal{S}_{\textrm{fact}}$, $\mathcal{S}_{\textrm{foil}}$. 
For unilateral explanations, only $\mathbf{y}_{\textrm{fact}}$ was provided to the LLM. Templates which guided the LLM to generate the explanations are provided in the Appendix~\ref{appx:prompts}. We used the OpenAI API~\cite{openai_api} and chose GPT-4 to generate the explanations~\footnote{The first author manually reviewed the generated explanations to verify whether the LLM introduced any hallucinations; we elaborate on this process in the limitations section.}.
\section{Experiment}

\subsection{Task description}

Participants were shown vignettes of fictitious characters and were asked to select the optimal exercise for the character in question based on their goals, capabilities, and preferences. They had to make a selection of the top exercise among 7 exercises, which were fixed choices across vignettes and alphabetically ordered in the drop-down list: aerobics, bicycling, boxing, jog/walk combination, pilates, resistance training, and swimming.

\subsection{Conditions}

Participants were randomized into one of the five conditions:\textit{no AI}, \textit{unilateral}, \textit{contrastive predicted}, \textit{contrastive after}, and \textit{contrastive random}, as described in Section~\ref{section:overview}. Figure~\ref{fig:exercise-task} provides a sample of a decision task with illustrations of the key conditions.

\subsection{Procedure}

Participants accessed the study online through Prolific, where they first provided informed consent. They then completed pre-task questionnaires, including a brief demographic survey, a six-item Need for Cognition (NFC) Scale~\cite{lins2020very}, and a seven-item Actively Open-minded Thinking (AOT) Scale~\cite{haran2013role}. The study consisted of three blocks: pre-test and post-test blocks, each with 5 exercise prescription tasks without AI support which served for measuring human learning, and an intervention block with 14 tasks where participants interacted with one of the AI interaction designs (or no AI, depending on their randomization). After completing the tasks, participants filled out a shortened version of the Intrinsic Motivation Inventory (IMI)~\cite{ryan1982control, mcauley1989psychometric}, a self-reported instrument intended to measure participants' subjective experience with the task, which assessed their perceived autonomy, competence, relatedness to AI, and interest/enjoyment, using 4 questions for each construct (except for relatedness, for which 3 questions where used). An additional question was included to assess mental demand.

\subsection{Participants}

We conducted a power analysis using G*Power~\cite{erdfelder1996gpower} to determine the required sample size for detecting a small effect size in our study with 5 conditions. With a small effect size, an $\alpha$ error probability of 0.05, and a desired power of 0.80, the analysis indicated that a total of 548 participants would be needed to achieve sufficient power to detect the effect. To account for filtering of spammers, a total of 800 participants were recruited to complete the task via Prolific. Participation was limited to US adults fluent in English. 
Recruited in batches, participants received an average compensation of \$2.70 (USD) per task. To ensure a compensation rate of \$12 per hour, we adjusted the payment from \$2.40 in the initial small batches to \$2.75 in later batches, reflecting the median time participants spent on the study. The average age of participants was $M = 35.76$ ($SD = 11.71$) and their education distribution was 0.5\% pre-high school, 19.4\% high school, 75.8\% college, 5.7\% post-graduate degree, and 4.6\% did not disclose their education.

\subsubsection{Exclusion criteria}

We retained 628 participants for analyses. To ensure meaningful engagement, participants with a median response time under 4 seconds were excluded, as this suggested insufficient consideration of the tasks, which required reading vignettes and selecting exercises. Those with any response time exceeding 2.5 minutes (90th percentile) were also removed to avoid data distortion from distractions. Additionally, participants in AI-assisted conditions who performed near random (below 20\% accuracy) or selected the same exercise for more than half of the study were excluded for potential misunderstanding. For subjective experience analyses, 6 participants were removed due to technical issues they encountered during the post-study questionnaire.

\subsection{Approval}

This study received approval from our institution's IRB under protocol number IRB21-0805.

\begin{figure*}[!h]
     \centering
     \begin{subfigure}{0.45\textwidth}
     \includegraphics[width=\textwidth]{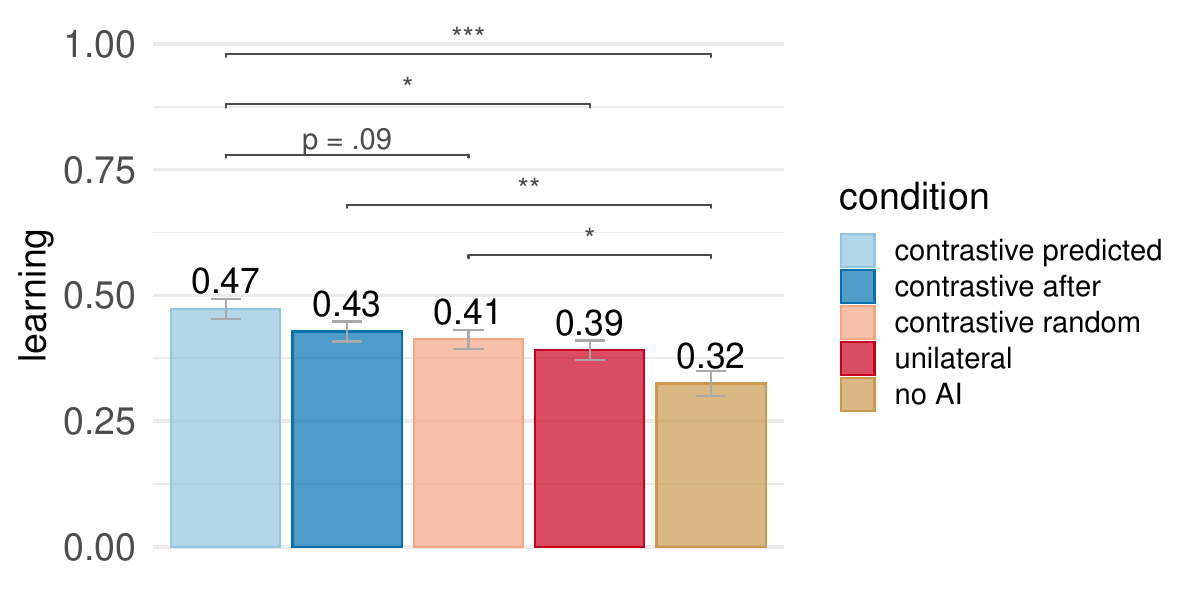}
     \caption{Learning}
     \label{fig:results-learning}
     \end{subfigure}
     \begin{subfigure}{0.45\textwidth}
     \includegraphics[width=\textwidth]{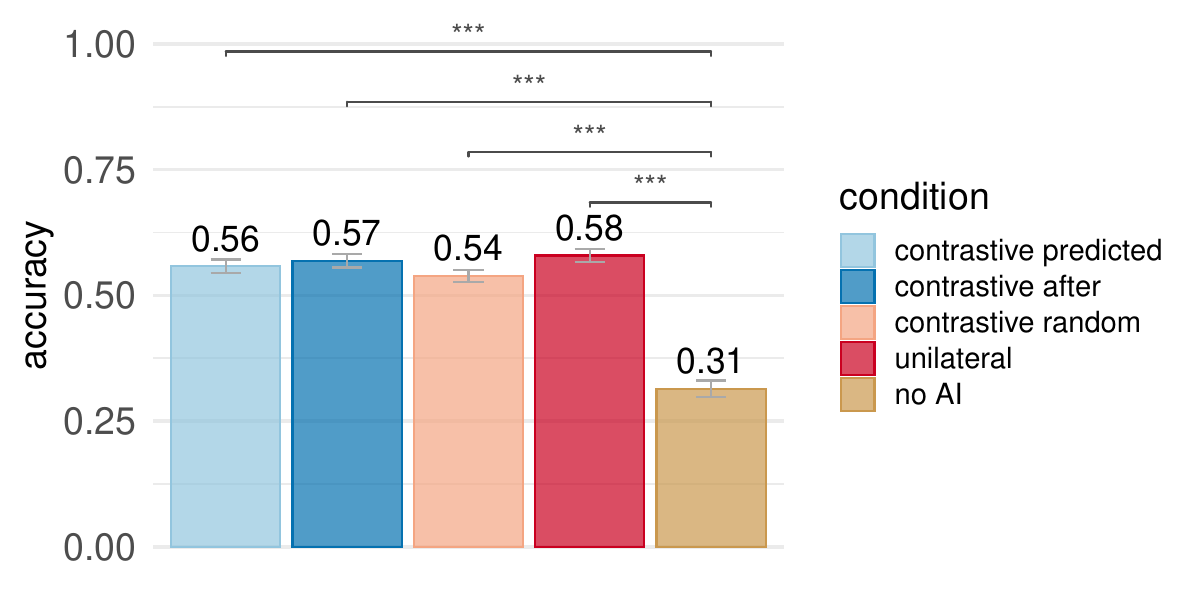}
     \caption{Accuracy}
     \label{fig:results-accuracy}
     \end{subfigure}
     \caption{Main results. Marginal means of human learning (post-intervention performance, controlled for pre-intervention performance) and accuracy accross different conditions. Error bars represent one standard error. Significance levels after Holm-Bonferroni correction are presented only for significant (or marginally significant) differences,  indicated by: * p < 0.05, ** p < 0.01, *** p < 0.001.}
\end{figure*}

\subsection{Design \& Analysis}

This study followed a between-subjects design, with the condition as the factor. Each participant interacted with one of the five conditions.

We collected the following indicators of performance and learning:

\begin{itemize}
    \item \textbf{Accuracy:} Percentage of correct answers provided by participants in the intervention block, where a correct answer is one that matches the ground truth. 
    
    \item \textbf{Overreliance:} Percentage of  answers that matched the AI's suggestions in questions for which participants received AI support and the AI's suggestion was incorrect.
    
    \item \textbf{Learning:} Percentage of correct answers on \textit{post}-intervention questions (controlled by participant's performance on \textit{pre}-intervention questions). 
\end{itemize}
For accuracy, learning, and overreliance in text and in figures we report the \textit{marginal means} produced by the regression models that included performance on pre-intervention questions as a covariate.

To assess the subjective experience, we collected the following measures assessed on a 5-point Likert scale, unless stated differently (See Appendix~\ref{post-study-questionnaire} for the questionnaire):

\begin{itemize}
    
    \item \textbf{Perceived Competence:} Four questions adapted from the Intrinsic Motivation Inventory (IMI) to measure participants' feelings of effectiveness and competence in the task.
    
    \item \textbf{Perceived Autonomy (Choice):} Four questions adapted from the IMI capturing the degree of autonomy and freedom participants felt in their decision-making.
    
    \item \textbf{Relatedness to AI:} Three questions adapted from the IMI measuring participants' sense of connection and trust in the AI.

    \item \textbf{Interest/Enjoyment:} Four questions adapted from the IMI to assess participants' interest and enjoyment during the study.
    
     \item \textbf{Mental Demand:} A single question, measuring the cognitive effort required by participants.
    
\end{itemize}

To assess the effects of experimental conditions on learning, accuracy, and subjective measures, we employed analysis of covariance (ANCOVA). For human learning, ANCOVA was applied to the average post-intervention correctness per participant, with pre-test performance as a covariate and condition as a fixed factor. A Shapiro-Wilk test was conducted on the residuals to check the normality assumption, which was not violated $(W = .993, p = .137)$. Holm-Bonferroni corrections~\cite{holm79:simple,shaffer95:multiple} were used to adjust for multiple comparisons across our eight hypotheses and planned analyses related to learning. Adjusted p-values are reported wherever a correction was applied. 
For accuracy, we again used ANCOVA, this time on the average correctness during the intervention. Pre-test performance was included as a covariate due to its significant correlation with intervention question performance, while condition was treated as a fixed factor.
Subjective measures were analyzed using ANOVA, with condition as the fixed factor. Post-hoc pairwise comparisons between conditions were corrected using Holm-Bonferroni method to account for multiple hypotheses.
Throughout the results, we report effect sizes using Cohen's $d$ along with 95\% confidence intervals. Effect sizes and accompanying confidence intervals provide valuable information, particularly when interpreting results where we hypothesize no significant differences. When the confidence interval for an effect size includes 0, it suggests that the true effect could be negligible or even nonexistent~\cite{colegrave2003confidence, lee2016alternatives, thompson2007effect}.  
For correlations, Pearson's $r$ is provided. 
\section{Results}

\subsection{Main results}

\subsubsection{Human Learning}

Main results for learning are depicted in Figure~\ref{fig:results-learning}. We report adjusted p-values, corrected with Holm-Bonferroni to account for multiple comparisons.
As hypothesized (\textbf{H-L1a} \& \textbf{H-L1b}), participants experienced statistically significantly greater learning in the \textit{contrastive predicted} ($M = 0.47$, $F_{1, 209} = 38.62$, $p = 0.00004$, $d = 0.65 \ [0.37, 0.94]$) and \textit{contrastive after} ($M = 0.43$, $F_{1, 216} = 26.68$, $p = 0.006$, $d = 0.47 \ [0.19, 0.74]$) conditions compared to participants in the no AI condition ($M = 0.32$).

Participants in the \textit{contrastive random} condition also showed significantly higher gains than those in the no AI condition ($M = 0.41$, $F_{1, 230} = 23.42$, $p = 0.02$, $d = 0.40 \ [0.13, 0.67]$). Conversely, participants in the \textit{unilateral} condition did not signficantly improve their learnring compared to \textit{no AI} ($M = 0.39$, $F_{1, 222} = 29.66$, $p = n.s.$, $d = 0.30 \ [0.03, 0.57]$).

Comparing contrastive conditions with sensible foil to unilateral explanations, as hypothesized (\textbf{H-L2a}), participants in the \textit{contrastive predicted} condition ($M = 0.47$) learned statistically significantly more than participants in the \textit{unilateral} condition ($M = 0.39$, $F_{1, 260} = 40.99$, $p = 0.02$, $d = 0.35 \ [0.11, 0.60]$). However, the difference between \textit{contrastive after} ($M = 0.43$) and \textit{unilateral} explanations was not significant ($F_{1, 267} = 33.05$, $p = n.s.$, $d = 0.16 \ [-0.08, 0.40]$), not lending support to \textbf{H-L2b}.

Within the contrastive conditions, participants in the \textit{contrastive predicted} condition demonstrated greater learning ($M = 0.47$) compared to those in the \textit{contrastive random} condition ($M = 0.41$). However, this difference was only marginally significant ($F_{1, 268} = 31.82$, $p = 0.09$, $d = 0.26 \ [0.02, 0.50]$), offering partial support for hypothesis \textbf{H-L3}. Addressing research question \textbf{RQ-L1}, the \textit{contrastive predicted} condition did not result in significantly different learning compared to the \textit{contrastive after} condition ($M = 0.43$, $F_{1, 254} = 41.44$, $p = n.s.$, $d = 0.18 \ [-0.07, 0.43]$).

\begin{figure}
     \centering
     \includegraphics[width=0.4\textwidth]{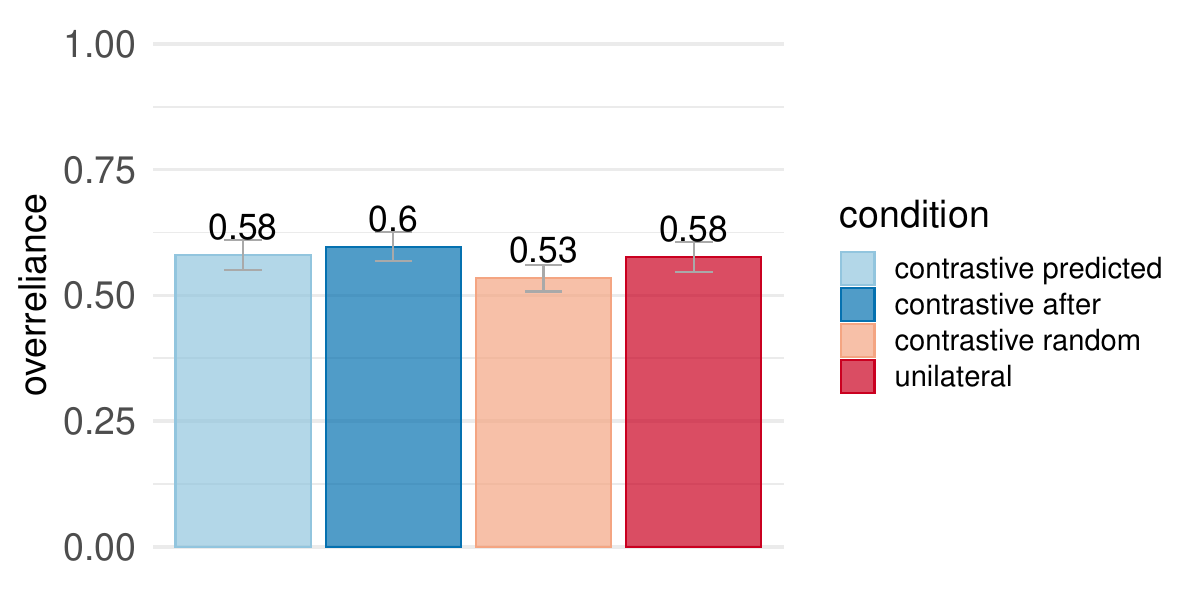}
     \label{fig:overreliance-by-condition}
  
     \caption{Overreliance across conditions. Error bars represent one standard error.}
\end{figure}

\subsubsection{Accuracy and Overreliance}

\begin{figure*}
     \centering
     \includegraphics[width=0.9\textwidth]{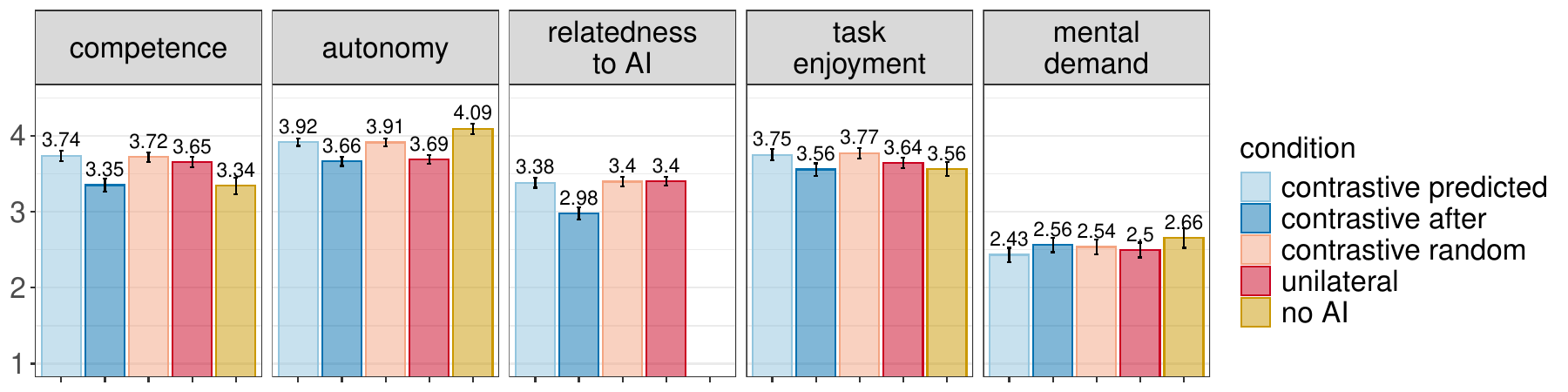}
     
     \caption{Subjective results. Error bars represent one standard error.}
     \label{fig:results-subjective}
\end{figure*}

Figure~\ref{fig:results-accuracy} summarizes results of human accuracy on the decision task with different conditions. 
As hypothesized (\textbf{H-A1a} \& \textbf{H-A1b}), the accuracy of participants in the \textit{contrastive predicted} ($M = 0.56$, $F_{1, 260} = 20.04$, $p = n.s.$, $d = -0.15 \ [-0.39, 0.10]$) and \textit{contrastive after} ($M = 0.57$, $F_{1, 267} = 9.13$, $p = n.s.$, $d = -0.08 \ [-0.32, 0.16]$) conditions was not significantly different from that of participants in the \textit{unilateral} condition ($M = 0.58$).

While participants improved their performance on the task significantly on average when they received AI support ($M = 0.56$) compared to receiving no AI support ($M = 0.31$, $F_{1, 627} = 195.32$, $p \ll 0.0001$), their performance also significantly degraded when AI suggestions were suboptimal ($M = 0.14$) compared to receiving no support ($M = 0.29$, $F_{1, 627} = 36.34$, $p \ll 0.0001$). Note that the different means for no AI support ($M = 0.31$, $M = 0.29$) in this analysis occur because we split the performance of participants in the no AI condition based on whether AI, if provided, would have been correct or incorrect, to allow a fairer comparison with other conditions that received incorrect suggestions for only a subset of questions. 

In situations when AI provided a suboptimal recommendation, participants in the \textit{contrastive predicted} ($M = 0.58$, $F_{1, 260} = 6.53$, $p = n.s.$, $d = 0.03 \ [-0.22, 0.27]$) and \textit{contrastive random} ($M = 0.54$, $F_{1, 281} = 3.83$, $p = n.s.$, $d = -0.12 \ [-0.35, 0.12]$) exhibited similar overreliance as those in the unilateral condition ($M = 0.58$), addressing \textbf{RQ-A1}.

Similarly, presenting contrastive explanations immediately (\textit{contrastive predicted}), resulted in similar overreliance ($M = 0.58$)  compared to presenting contrastive explanations after a decision was made (\textit{contrastive after}) ($M = 0.59$, $F_{1, 254} = 10.61$, $p = n.s.$, $d = -0.01 \ [-0.25, 0.24]$). 


\subsubsection{Subjective Experience}

Subjective results are summarized in Figure~\ref{fig:results-subjective}.


Condition was a significant predictor of perceived competence ($F_{4, 618}=6.40$, $p \ll 0.00005$). A Holm-Bonferroni corrected post-hoc test revealed that participants in the \emph{contrastive predicted}, \emph{contrastive random}, and \emph{unilateral} conditions reported significantly higher competence compared to those in the \emph{contrastive after} and \emph{no AI} conditions.

Perceived autonomy (\emph{i.e.}, choice)  was also significantly predicted by condition ($F_{4, 618}=8.85$, $p \ll 0.00001$).
A Holm-Bonferroni corrected post-hoc test revealed that participants in the \emph{contrastive predicted}, \emph{contrastive random}, and \emph{no AI} conditions perceived significantly higher autonomy compared to those in the \emph{unilateral} and \emph{contrastive after} conditions.


Condition was also a significant predictor of relatedness to AI (computed only for conditions involving AI) ($F_{3, 533}=9.02$, $p \ll 0.00001$), with a Holm-Bonferroni post-hoc test revealing that participants in the \emph{contrastive after} condition felt significantly less related to the AI compared to those in other AI conditions.

Task enjoyment/interest was not significantly predicted by condition ($F_{4, 618}=1.66$, $p = n.s.$) and neither was mental demand ($F_{4, 618}=0.47$, $p = n.s.$).

Across measures, the subjective results support \textbf{H-S2}, showing that contrastive explanations with a predicted foil led to significantly higher perceptions of competence, autonomy, and relatedness to the AI compared to the contrastive after condition. Additionally, our findings partially support \textbf{H-S1}: while contrastive explanations with a predicted foil significantly increased perceived autonomy compared to unilateral explanations, no significant differences were observed for competence and relatedness to the AI. 


\subsubsection {Subjective vs. objective measures}

\begin{figure*}
     \centering
     \includegraphics[width=0.7\textwidth]{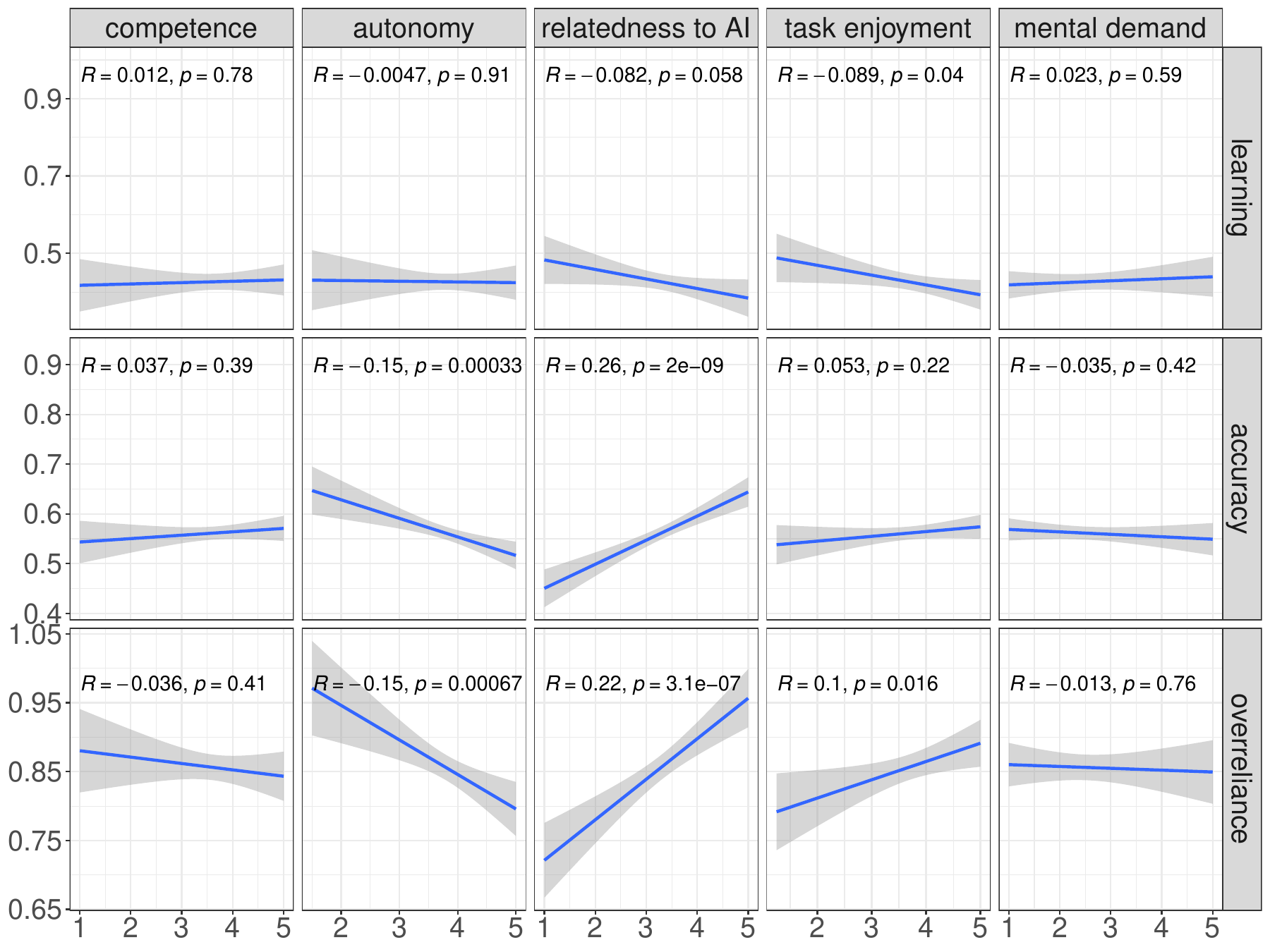}
     
     \caption{Relationship between subjective experience vs. objective outcomes. R indicates Pearson's $r$. Only conditions with AI support are included in the analysis.}
     \label{fig:subjective-vs-objective}
\end{figure*}

Figure~\ref{fig:subjective-vs-objective} shows the relationship between subjective experience and objective outcomes across conditions with AI support (\em i.e. \em, no AI condition was not included in the analysis). Our analysis revealed that there was no correlation between actual learning and competence, autonomy, or mental demand and that actual learning was very weakly inversely correlated with relatedness to AI ($r = - 0.08, p = 0.06$) and task enjoyment ($r = - 0.09, p = 0.04$).
Accuracy was significantly positively correlated with relatedness to AI ($r = 0.26, p \ll 0.0001$), a construct that included questions about trust in AI too, and it was significantly negatively correlated with perceived autonomy ($r = -0.15, p = 0.0003$). Similarly, overreliance was significantly positively correlated with relatedness to AI ($r = 0.22, p \ll 0.0001$), and significantly negatively correlated with perceived autonomy ($r = -0.15, p = 0.0006$). In addition, overreliance was significantly positively correlated with task enjoyment ($r = 0.1, p = 0.02$).




\subsection {Audit for intervention-generated inequalities}

\begin{figure*}
     \centering
     \begin{subfigure}{0.48\textwidth}
     \includegraphics[width=\textwidth]{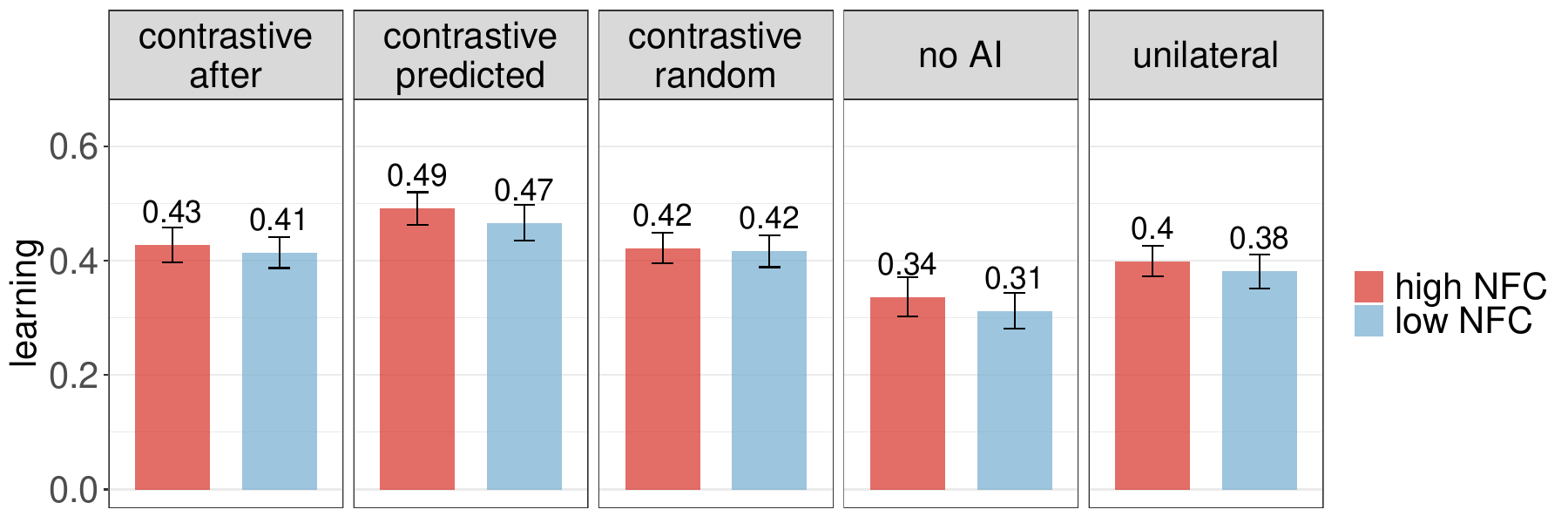}
     \caption{Need for Cognition (NFC)}
     \label{fig:learning-NFC}
     \end{subfigure}
     \begin{subfigure}{0.48\textwidth}
     \includegraphics[width=\textwidth]{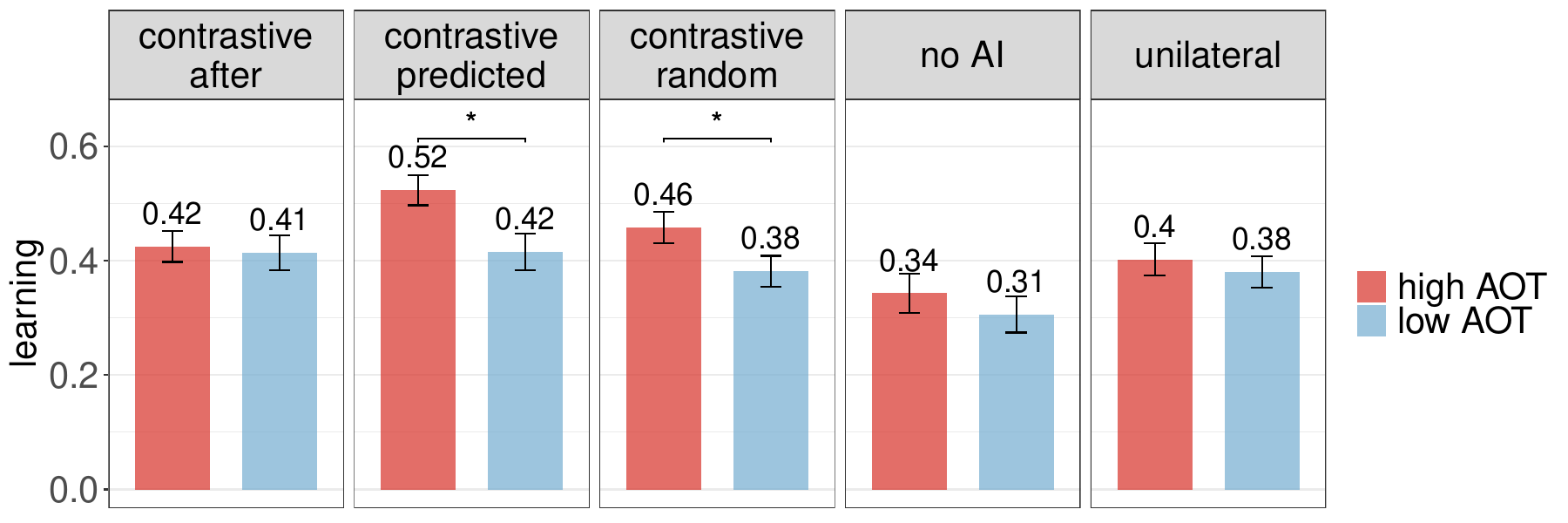}
     \caption{Actively Open-minded Thinking (AOT)}
     \label{fig:learning-AOT}
     \end{subfigure}
     \caption{Auditing for intervention generated inequalities: Learning (marginal means) for different individual differences. Error bars represent one standard error. Significance levels after Holm-Bonferroni correction are indicated by: * p < 0.05.}
\end{figure*}

Intervention-generated inequalities occur when an intervention, while beneficial on average, disproportionately benefits some groups over others~\cite{lorenc2013types}. Disaggregating results by relevant demographics or variables can help uncover these disparities. Informed by prior research in AI-assisted decision-making~\cite{bucinca2021trust, gajos2022people}, we conduct a self-audit and examine whether contrastive explanations, introduced as interventions to enhance human decision-making skills, benefit different groups equally.

Previous studies have shown that individual differences in information processing can significantly impact the effectiveness of AI support and interventions, particularly for cognitively demanding outcomes like learning. One individual difference that may affect the effectiveness of our interventions is Need for Cognition (NFC), a stable trait that reflects an individual's motivation to engage in deep thinking and information processing~\cite{cacioppo1982need}. NFC has been consistently identified as a predictor of performance in cognitive tasks such as problem-solving and decision-making~\cite{cacioppo96:dispositional}. In the context of AI-assisted decision-making, NFC has been found to influence whether cognitive forcing reduces overreliance on AI~\cite{buccinca2021trust} and how effectively individuals learn from AI assistance~\cite{gajos2022people, buccinca2024towards}.

Another important individual difference that we reasoned would be particularly relevant for interventions that require consideration of multiple viewpoints is Actively Open-Minded Thinking (AOT). People high in AOT are more likely to critically evaluate new evidence, weigh it against their existing beliefs, take sufficient time to solve problems, and carefully consider others' opinions when forming their own~\cite{baron1993teach, haran2013role}. We investigate whether individuals with varying levels of AOT benefit differently from contrastive explanations, which provide alternative ``viewpoints'' for consideration.

Figures~\ref{fig:learning-NFC} and ~\ref{fig:learning-AOT} depict results disaggregated by NFC and AOT. We did not find any significant differences among the effectiveness of (any) contrastive explanations for people with different levels of NFC (for detailed ananlyses see Appendix, Table~\ref{tab:nfc-results}). 
However, our findings reveal a notable contrast in the AOT groups: participants with high AOT benefited significantly more from the \emph{contrastive predicted} condition ($M=0.52$) compared to those with low AOT ($M=0.42$; $F_{1, 122} = 6.67$, $p = 0.01$, $d = 0.47$ [$0.11$, $0.84$]). Similarly, the \emph{contrastive random} condition was more effective for individuals with high AOT ($M=0.46$) than for those with low AOT ($M=0.42$; $F_{1, 142} = 3.77$, $p = 0.05$, $d = 0.34$ [$-0.01$, $0.68$]). See Table~\ref{tab:aot-results} for non-significant conditions.
These findings uncover AOT as a relevant individual difference to consider in AI-assisted decision-making and reveal that contrastive explanations may unevenly impact individuals, offering greater advantages to those with higher AOT.

\section{Discussion}

We investigated whether AI decision support systems that account for the decision-maker's mental model of the task and explain their misconceptions can simultaneously enhance decision accuracy and promote the development of independent decision-making skills. 

\subsection{On the effectiveness of contrastive explanations in improving human-AI decision-making outcomes}

As expected, our results showed that participants learned significantly more with contrastive explanations with predicted foil compared to unilateral explanations (\textbf{H-L2a}) or no AI support (\textbf{H-L1a}) (Figure~\ref{fig:results-learning}). Moreover, also as hypothesized (\textbf{H-A1a}), this improvement in learning was achieved without sacrificing accuracy: participants completing the task with contrastive explanations with predicted foil were as accurate as their counterparts who received unilateral explanations (Figure~\ref{fig:results-accuracy}). Additionally, participants in the contrastive explanations with predicted foil condition reported significantly greater perceived autonomy (but not competence or relatedness to AI) during the task compared to those in the unilateral condition, providing partial support for \textbf{HS-1}. 

The \emph{contrastive after} condition, where participants received contrastive explanations after making an initial decision (inputted foil), led to significant learning gains compared to receiving no AI support (lending support to \textbf{H-L1}) but not significantly different learning compared to unilateral explanations (not supporting \textbf{H-L2b}). As expected (\textbf{H-A1b}), participants' accuracy in the contrastive after condition was not significantly different from those in the unilateral condition. 

Overall, our research provides compelling evidence that contrastive explanations with predicted foils significantly enhance decision-making skills without sacrificing decision accuracy compared to unilateral explanations, which remain the default explanation design in AI-powered decision support. Our study is the first to demonstrate that even when AI offers decision recommendations (rather than explanations alone~\cite{gajos2022people, buccinca2024towards}), users can still cognitively engage with its content and improve their learning about the task when this content is engaging. This finding opens new possibilities for optimizing AI decision-support systems by intervening not only at the \emph{interaction} level, as previous work suggests~\cite{buccinca2021trust, buccinca2024towards, buccinca2022beyond, zhang2024beyond}, but also at the \emph{content} level of the explanations themselves to improve human-AI decision-making outcomes. 

Lastly, while contrastive explanations with predicted foil improved human decision-making skills, we do not think they are a panacea for human-AI decision-making. For example, our results showed that contrastive explanations (as well as unilateral explanations) still resulted in significant overreliance on AI. Also, they were signficantly more effective for people high in AOT (who are inherently driven to consider multiple viewpoints) compared to those low in AOT.  Instead, we believe that contrastive explanations are useful when shown in the right situations, such as when the AI is confident about its decision, and to people who benefit from them (e.g., those high in AOT). As such, these explanations expand the suite of human-AI interaction techniques that can be adaptively selected in appropriate situations to optimize human-AI decision-making outcomes, like decision accuracy and human learning~\cite{buccinca2024towards, swaroop2024accuracy, bhatt2023learning}. 

\subsection{What have we learned about the design of contrastive explanations?}

Our results provide evidence about which aspects of contrastive explanations matter for objective and subjective outcomes in human-AI decision making. 

First, as hypothesized (\textbf{H-S2}), our findings show that interaction design matters for subjective experience: contrastive explanations are as effective in objective measures when the foil is predicted as they are when the foil is inputted (the contrastive after condition) --- even though the inputted foil is the ``perfect'' comparison. However, consistent with prior research~\cite{bucinca2021trust, fogliato2022goes}, our results show that providing contrastive explanations \emph{after} people make their own decisions (input their foil) results in significantly lower subjective experience, even if that advice engages with their own input as in contrastive explanations after condition. We found no differences in subjective experience between contrastive explanations with a predicted foil and those with a random foil, suggesting that the contrastive design, applied before a decision is made, is perceived favorably regardless of the foil's quality.

Second, our results show suggestive evidence that quality of the foil matters for improving learning as the objective outcome of the interaction. When contrastive explanations are presented at the decision-making time, high quality foil such as in \emph{contrastive predicted} resulted in greater learning on average compared to a randomly selected foil, albeit the difference was only marginally significant, partially supporting \textbf{HL-3}. We believe that one of the reasons why the difference between these two conditions is not more pronounced in our study is that even a ``random'' foil in our setting is relatively reasonable. A randomly selected exercise from the list still addresses at least part of the needs or preferences of the fictitious character, rendering it a choice worth considering. We believe that in different situations, such as medical treatment decisions, where the choices may consist of a wide variety of treatments for a wide array of diseases, a randomly selected choice would likely be obviously ineffective or harmful, thus a waste of cognitive resources for the clinician to consider. In addition, we believe there is room to further improve the quality of the predicted foil. In our implementation, the foil was generated using a single model that predicted the average human response across all decision-makers. We believe that employing personalized models, which capture each individual's unique mental model of the task, could result in even more accurate foils and, consequently, lead to greater learning gains. Our analysis of the participants' responses used to train the human model revealed high variability in exercise choices across participants (Appendix~\ref{appx:human-model-evaluation}), further supporting the need for personalized models. Future research should investigate how to best fine-tune models to individuals and assess the added value of personalized models compared to average human models for enhancing downstream human-AI decision outcomes.

In this study, we sought to deepen our understanding of the timing of contrastive explanations and the impact of foil quality. We experimented with a simple, intuitive design in which the foil represented the choice of many people, while the fact reflected the AI's suggestion in the user interface. However, contrastive reasoning can be conveyed in various other forms. For example, two conversational agents—one advocating for the human model's choice and the other for the AI's—could engage in a dialogue, allowing the decision-maker to assess which agent's reasoning is more compelling. Alternatively, designs could focus solely on contrastive dimensions, rather than the fact and foil, by highlighting aspects of the decision that the human decision-maker may be overlooking. This approach could provide insights as intermediate support without offering a direct recommendation (e.g., \textit{cardio supports weight loss goals}). Having demonstrated the effectiveness of one human-AI contrastive design in promoting learning, we believe future research should explore a wider range of design possibilities for representing human-AI misalignment in even more impactful ways.


\subsection{What have we learned about the effects of contrastive explanations on overreliance?}

Evidence from prior work suggested that presenting more than one AI suggestion (i.e., a ``second opinion'') to people may reduce their overreliance, as it makes them more likely to consider alternatives~\cite{lu2024does, bansal2021does}.  Our results showed that participants in contrastive explanations conditions (with predicted or random foils) exhibited similar rates of overreliance on AI suggestions as those in unilateral condition (\textbf{RQ-A1}). We believe that we may not be observing the beneficial effect of second opinion in our study because in situations when the simulated AI provided incorrect recommendations (i.e., when the ``fact'' was a suboptimal choice), the foil was an even worse choice. Therefore, participants were primed to contrast two suboptimal choices and resorted to the better choice out of the two. Future work should explore whether contrastive explanations would still result in similar overreliance, when the foil is a better alternative than the fact.

Interestingly, we also found that participants' overreliance rate on AI in contrastive after condition was similar to that of the unilateral condition. This contrasts with prior research showing that providing \emph{unilateral} explanations after an initial decision reduces overreliance~\cite{buccinca2021trust, green2019principles}, as people are less likely to follow incorrect AI advice once they have made a decision. In our study, because contrastive explanations directly addressed participants' decision and provided evidence as to why their choice was inferior to the AI's, they seemed more persuasive, potentially diminishing the positive effect of the cognitive forcing.


\subsection{What have we learned about intrinsic motivation in AI-assisted decision-making?}

We measured participants' perceived competence, autonomy, and relatedness \emph{to AI} as psychological needs underpinning individuals' intrinsic motivation about a task. Our results demonstrate that both interaction and explanation design significantly impact these constructs. First, as hypothesized \textbf{H-S2}, we found that the contrastive after condition---in which contrastive explanations ``critiqued'' individuals' inputted answer and presented evidence that AI's choice was superior---led to significantly lower perceived competence, autonomy and relatedness to AI compared to situations in which contrastive explanations were presented before a decision was made. 
Second, our results demonstrated that contrastive explanations provided before a decision (whether using a predicted or random foil), which presented two decision choices, led to significantly higher perceived autonomy in task completion---comparable to participants who received no AI support---compared to unilateral explanations that offered only a single option.

Our analysis of intrinsic motivation constructs and objective outcomes revealed that actual learning was not correlated with perceived competence. Increased perceived autonomy was correlated with reduced overreliance but also with lowered accuracy, while stronger perceptions of relatedness to the AI were correlated with greater overreliance on AI and higher accuracy.

These findings suggest that the design of AI support can significantly influence people's intrinsic motivation toward a task, as well as objective outcomes such as accuracy and overreliance. We believe that when developing new AI-assisted decision-making systems, researchers should carefully consider and measure how these designs affect people's intrinsic motivation about the task in addition to the objective outcomes of the interaction.

\subsection{Generalizability \& Limitations}

We conducted a single controlled experiment with a single task and with crowdworkers. While prior research on AI-assisted decision-making suggests that experts often exhibit similar behavior to non-experts when relying on AI systems~\cite{gaube2021ai}, we do not know whether this holds for learning from the AI about a task of their expertise. \citet{jacobs2021designing} show that clinicians would prefer a system that explains why AI's choice differed from the established clinical guidelines, which suggests they may be open for learning from the AI.
Our task choice had inherent learning opportunities (e.g., facts about exercises). Learning may not be as pronounced in some tasks, such as hiring, were opportunities for learning exist but are sparser. Moreover, we assessed only the short-term effects of the explanation designs on learning. The long-term benefits and whether the observed learning gains will persist over time remain unstudied.
Further research is needed to understand how generalizeable our findings are for other tasks, domains, and settings.

We believe that our contrastive explanations framework can be effectively applied to a wide range of tasks and settings. Its modular design allows for flexible adaptation based on specific contexts. What we called an expert model, in a new domain can be replaced by the predictive model that drives the decision recommendations. More flexibility exists in how the human model is constructed to generate sensible foils. If data exist on how human decision-makers made decisions in a domain, those data can be used to train the human model. In domains with established guidelines–where AI is introduced to enable more nuanced decisions than were previously possible --- the foil could be derived from those guidelines, enabling a comparison between established practices and AI-based predictions. Alternatively, the foil could reflect a personalized human model, such as one trained on an individual clinician’s past choices.
The other modules can also be customized according to the domain. For instance, the contrast module could compute pixel gradients or concept activations~\cite{kim2018interpretability} that highlight differences between the fact and foil. Similarly, the presentation module can be customized to suit the task, such as employing tailored visualization techniques. However, we believe that contrastive explanations---and by extension, our framework---are most valuable in multiclass classification or ranking scenarios, where the foil is less obvious than in binary decision contexts.

An important consideration about our work is that we chose to implement the presentation module with a large language model (LLM). We used the LLM to turn the scaffold produced by the rest of the modules into a natural language explanation, while providing small gaps in the template for it to fill with domain facts. We believe the approach of constraining the generation of facts within the constraints of more trusted predictive models may be useful for certain settings, such as ours but may not generalize to expert-level domains where the LLM may not have the nuance to fill in the gaps. 
Moreover, we iteratively arrived at prompts (included in the Appendix) which produced explanations with almost no hallucinations. The first author of the paper reviewed the generations for all the characters included in the experiment, finding the LLMs to only mix the indoor/outdoor preferences of characters at times, but no other major hallucinations. However, such manual review cannot be scaled. We believe that our framework can be extended to include a verification step for the generated explanations. For example, multiagent frameworks~\cite{wu2023autogen} can be used with additional agents reviewing the generated explanations.

Another limitation of our study is that, in order to control the AI’s mistakes, we chose to simulate the AI. We introduced errors in four randomly selected questions during the intervention phase, where the AI’s suggestion was generated by a human model rather than the expert model. This approach may have contributed to overreliance on the AI, as the wrong AI suggestion was a likely human choice.

Finally, in this study, we investigated the impact of explanation design on the learning of factual information. While our study was intentionally structured so that the AI consistently supported its reasoning with truthful facts, there remains a potential, yet unstudied (to the best of our knowledge), risk of inadvertent outcomes or misuse of designs that support learning. This risk could manifest as the propagation of misinformation if the AI were to provide incorrect explanations. To address this concern, we strongly advocate for the use of contrastive explanations --- or any form of explanations --- in conjunction with an intelligent interaction layer (e.g., as in ~\cite{buccinca2024towards, swaroop2025personalising, noti2022learning}). Such a layer would ensure that explanations are delivered exclusively when the predictive model demonstrates high confidence in its recommendations and reasoning, thereby reducing the likelihood of overreliance and minimizing the risk of misinformation.

\section{Conclusion}

In this work, we investigated whether explanation designs that account for human reasoning can improve human decision-making skills in the task in AI-assisted decision-making. We introduced a framework for generating human-centric contrastive explanations by showing the difference between AI's reasoning and a likely human response for the same task. Our results demonstrated that contrastive explanations significantly enhanced human decision-making skills compared to unilateral explanations, the default method of AI support, without compromising accuracy. Sparking hope about growing deskilling concerns, our work suggests that AI support that accounts for human mental models of the task can be a promising approach toward systems that augment and upskill decision-makers.

\begin{acks}
This work was supported in part by the NSF under Grant No. IIS-2107391 and ONR under Agreement No. N00014-24-1-2726. Any opinions, findings, and conclusions or recommendations expressed in this material are those of the author(s) and do not necessarily reflect the views of NSF or the views of ONR. We thank Christina Xiao for insightful discussions on contrastive explanations, Alexandra Chouldechova and Jenn Wortman Vaughan for valuable suggestions in selecting representative exercises for the interface, Eura Shin, Lucia Gordon, and Sohini Upadhyay for their feedback on the manuscript, and Paula Rodriguez-Diaz for expertly recalibrating the lab's coffee machine.
\end{acks}

\bibliographystyle{ACM-Reference-Format}
\bibliography{kzg}

\onecolumn
\appendix
\section{Appendix}

\subsection{Task Design: Implementation details}

\subsubsection{Evaluating the expert model}
\label{appx:evaluating-exper-model}

We developed the objective function and expert model iteratively over multiple discussion sessions with the expert. In each session, we evaluated the critical dimensions for inclusion in the model and assessed its predictions, deciding whether to add or remove dimensions accordingly. Once we had decided the structure of the $\mathbf{g}$ representation, the expert provided a total of 322 pairwise comparisons among exercises for 12 unique fictitious characters. For learning the expert weights for the objective function $\mathbf{f}$ in Equation~\ref{eq:objective-function}, we followed the approach described in Section~\ref{subsec:learning-expert-weights}. Using the Scikit-learn library in Python, we trained a Support Vector Machine (SVM) model with a linear kernel and a regularization parameter (C) set to 1.0.
We evaluated the model's performance using 12-fold cross-validation, where each fold excluded one of the fictitious characters. The model was trained on the remaining characters and tested on its ability to predict pairwise comparisons for the excluded character. The model achieved a mean accuracy of 0.86 with a standard deviation of 0.08 across all folds. Additionally, the mean area under the ROC curve (AUC) was 0.86. It is important to note that these results reflect pairwise comparisons involving the full set of 59 exercises, and not only the subset of 7 exercises with which we populated the drop-down list in the interface.
For the final step, we qualitatively assessed the model's choices for new fictitious characters, confirming that the decisions were sound and reasonable. (Providing additional validation that the model effectively captures expert reasoning about the designed task, another self-identified kinesiology expert, who participated in one of our formative studies online, achieved a 96\% score in the task---significantly higher than the typical ~32\% average from crowds.)

\subsubsection{Selecting the drop-down exercises}
\label{appx:selecting-drop-down-exercises}

\begin{figure*}
     \centering
     \includegraphics[width=\textwidth]{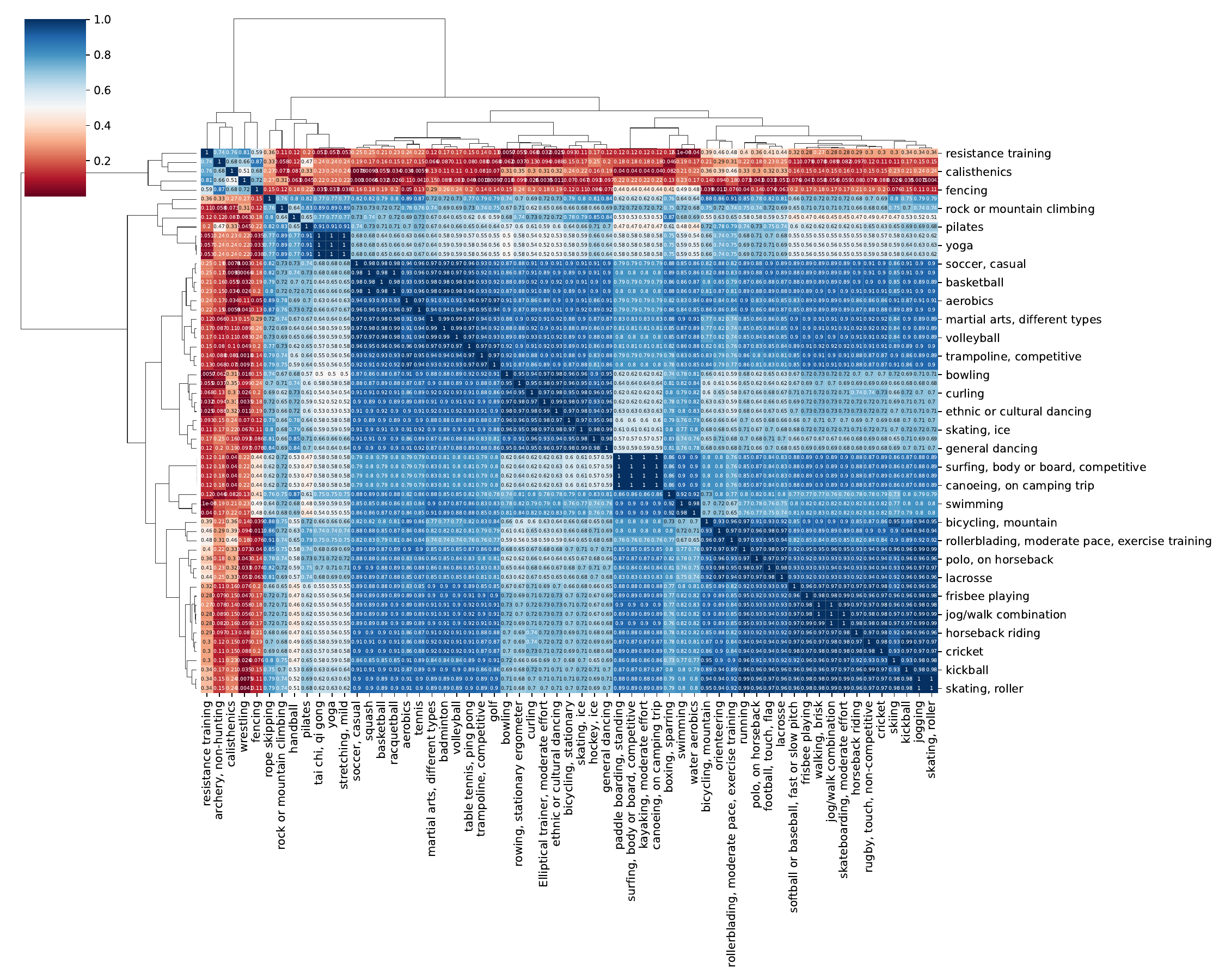}
     \caption{Correlation heatmaps with hierarchical clustering that served as a basis to select the drop-down exercises for the interface.}
     \label{fig:hierarchical-clustering}
\end{figure*}

We sought to populate the drop-down list for the interface with a sensible number of exercises that would not overwhelm the participants. To select a representative set of exercises from the larger set of the 59 exercises, we clustered the exercises based on their similarity. We generated a large set of 300 fictitious characters, and scored each of the 59 exercises for each of the characters with the expert scoring function. We then computed the correlations between the scores of the exercises and clustered them based on the similarity of their score profiles using hierarchical clustering (as depicted in Figure~\ref{fig:hierarchical-clustering}. This method allowed us to group exercises that received similar scores across the 300 characters into clusters. We applied agglomerative clustering with Ward’s method. After generating the dendrogram, we determined an appropriate number of clusters by examining the level at which the clusters remained distinct while minimizing redundancy across exercises.

To select representative exercises from each cluster, we calculated the centroid of each cluster, representing the average score profile across all exercises in that group. From there, we selected the exercise whose score profile was closest to the centroid and that was also a more common or accessible exercise (e.g., aerobics vs. trampoline jumping) , ensuring that the selected exercise would be a good representative of the group as a whole. A set of 7 representative exercises (\emph{aerobics}, \emph{bicycling}, \emph{boxing}, \emph{jog/walk combination}, \emph{pilates}, \emph{resistance training}, \emph{swimming}) was then used to populate the drop-down list in the interface, providing a diverse but manageable selection that reflected the range of exercise options without overwhelming participants with too many choices.

\subsection{The Contrastive Explanation Framework: Implementation Details}
\subsubsection{Data collection study for training the human model}
\label{appx:data-collection}
We conducted an online user study on Prolific for collecting data with which to train the human model. The task and procedure were identical to those used in the main experiment, but participants completed the task without AI assistance, as our goal was to capture the human mental model of the task. In total, 20 participants answered 100 questions, with each participant selecting exercises for 5 characters, randomly sampled from 30 unique characters (distinct from the characters used in the main experiment). Participants achieved a mean accuracy of ~30\% on the task. 

\begin{figure*}
     \centering
     \includegraphics[width=0.5\textwidth]{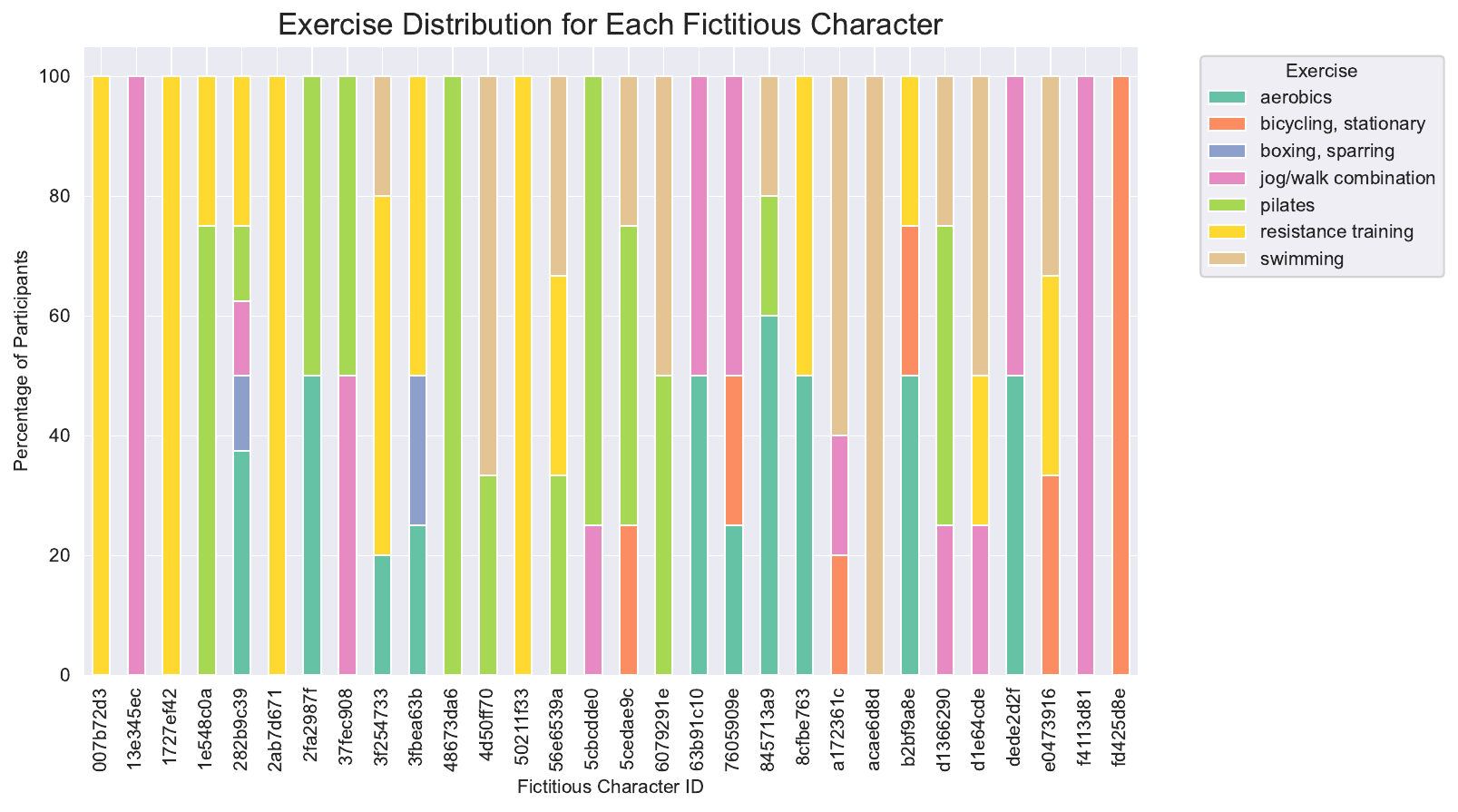}
     \caption{Distribution of participants' responses for the 30 fictitious character in the data collection study.}
     \label{fig:distributions-of-participant-responses-per-character}
\end{figure*}

Figure~\ref{fig:distributions-of-participant-responses-per-character} shows the distributions of exercise choices participants selected per fictitious character. To evaluate the variability of participants' responses in selecting exercises for different fictitious characters, we computed normalized entropy~\cite{shannon1948mathematical} per fictitious character. High variability could signal differing decision strategies for the task, while low variability would indicate stronger consensus and shared mental model. We selected \textit{normalized entropy} as it provides a robust measure of uncertainty, independent of the number of available categories, making it ideal for comparing variability across different characters. With a computed mean normalized entropy of $\mu = 0.51$, we found that participants' choices exhibited moderate variability, indicating that while some patterns emerged, responses remained fairly distributed across exercise choices. The standard deviation of $\sigma = 0.35$ further showed notable fluctuation in variability across characters, implying that certain fictitious characters elicited more consistent responses, whereas others triggered more diverse decision-making.
This analysis informed our evaluation of the human model, as we expected a moderately, but not highly, accurate model given the variability of participants' responses. 


\subsubsection{Evaluating the human model}
\label{appx:human-model-evaluation}

We trained the human model from the responses collected in the data collection study and by following the same approach with which we learned the expert weights. The total 100 choices (each out of 7 exercises) from the data collection study, yielded a total of 600 pairwise exercise comparisons. As with the expert model, we trained a Support Vector Machine (SVM) model with a linear kernel and a regularization parameter (C) set to 1.0. We used a 30-fold cross-validation for evaluating the human model, where in each fold one participant was removed from the training set. The model was then trained on the remaining participants and tested on the excluded one. This process was repeated for all participants, allowing us to assess the model's generalizability and its ability to predict individual behavior across different subsets of the data. As expected from the high variance in participant responses, the model was moderately accurate, with a mean cross-validation accuracy of 0.69 and an AUC of 0.68 for the pairwise comparisons of exercises.

As an additional evaluation, we compared the human model to the expert model. We generated a new set of characters to conduct the evaluation and found that 60\% of the unseen 50 characters, both human and expert models produced the same responses. The key differences emerged in specific exercise choices: the human model was less likely to select \emph{boxing} or \emph{aerobics}, which the expert model identified as suitable for some characters. Despite the high variability within and across participants, this demonstrates a ``wisdom of the crowd'' effect~\cite{surowiecki2005wisdom}, where the average human model captured signal across participants' responses (achieved 60\% accuracy on the task, compared to average participant accuracy of 30\%), resembling the expert model (an effect also observed in~\cite{grgic2024denoise}). In the main study, for cases where the human and expert models provided identical responses for the characters, we selected the second-highest-ranked option from the human model as the human response (i.e., the foil).

Finally, we sought to compare the outputs of our human-centered foil generator with those of the approach commonly used in prior work, which typically selects the second-best alternative from the model's perspective. To address this, we conducted an additional analysis. Specifically, we generated foils for 10 datasets, each containing 100 fictitious characters. We computed how often the human-centered foil selected by our approach matched the foil generated by the model-centered approach (i.e., the second-best option according to the model). The mean agreement was $M=0.627, 95\% CI [0.60, 0.65]$, indicating that our approach selects a different foil approximately 35\% to 40\% of the time, highlighting a notable divergence from model-centered foil selection methods.

\subsection{LLM Prompts}
\label{appx:prompts}
The variables in double brackets were populated according to the character in question.

\subsubsection{Contrastive Explanation Prompt}

\begin{quote}
\texttt{[[vignette]]}

Here are the aspects that a kinesiology expert considers when making the decision:
\begin{enumerate}
    \item Intensity: whether the intensity required to carry out an exercise exceeds the fitness capabilities of the person.
    \item Intensity: whether an exercise matches the intensity the person is capable of exerting.
    \item Goal: whether the exercise matches the person's goals.
    \item Preference: whether the exercise matches the person's preference.
\end{enumerate}

According to the expert's function, \texttt{[[fact]]} is better than \texttt{[[foil]]} on the following: \texttt{[[positive\_contributors\_fact]]}. Whereas, \texttt{[[foil]]} is better than \texttt{[[fact]]} because of: \texttt{[[positive\_contributors\_foil]]}.

Construct an explanation about why \texttt{[[fact]]} is better than \texttt{[[foil]]} using the following template:

Make it compact. Go into bullet point(s) strictly only for concepts for which the fact is better than the foil according to the expert's function. Do not explicitly say anything about the expert. Acknowledge the benefits of the foil over the fact if any as the first sentence, then highlight the tradeoffs in high-level concepts at the beginning of the explanation. 

Use the following structure for each bullet point, one by one, and include only the concepts for which \texttt{[[fact]]} is superior to \texttt{[[foil]]}:

\begin{itemize}
    \item Identify the primary characteristic of the superior exercise (e.g., running is a cardio exercise) and contrast this to the other exercise.
    \item Connect this characteristic to a benefit relevant to the character (e.g., cardio is good for weight loss).
\end{itemize}

High-level sentence that first acknowledges the concepts for which the foil is better than the fact (if any) or states that the foil is also a good choice, then highlights the concepts for which the fact is superior to the foil. Include only the concepts for which \texttt{[[fact]]} is superior to \texttt{[[foil]]}.

\begin{itemize}
    \item Concept 1 (e.g., Goal): 
    \item Concept 2: ...
\end{itemize}

Format the response as a JSON object:\\{"high\_level\_contrastive\_explanation": "explanation", "contrast\_concepts": [\{"Formatted name of concept (e.g., Goal)": "explanation"\}]}.
\end{quote}

\subsubsection{Unilateral Explanation Prompt.}

\begin{quote}
\texttt{[[vignette]]}

Here are the aspects that a kinesiology expert considers when making the decision:
\begin{enumerate}
    \item Intensity: whether the intensity required to carry out an exercise exceeds the fitness capabilities of the person.
    \item Intensity: whether an exercise matches the intensity the person is capable of exerting.
    \item Goal: whether the exercise matches the person's goals.
    \item Preference: whether the exercise matches the person's preference.
\end{enumerate}

Create a concise explanation for why \texttt{[[fact]]} is the best exercise for the specified character, using the following structure in bullet points only for aspects the expert considers:

\begin{itemize}
    \item Identify the primary characteristic of \texttt{[[fact]]} (e.g., running is a cardio exercise).
    \item Connect this characteristic to a benefit relevant to the character (e.g., cardio is beneficial for weight loss).
\end{itemize}

Strictly only include aspects recognized by the expert as beneficial for the character, omitting any for which \texttt{[[fact]]} may not be optimal or relevant. Do not explicitly say anything about the expert. Use the terms \texttt{Goal}, \texttt{Intensity}, and \texttt{Preference} when describing the relevant 'concept'.

Format the response as a list of JSON records with \texttt{'concept'} and \texttt{'explanation'} as the keys for the records.
\end{quote}

\subsection{Post-study Questionnaire}
\label{post-study-questionnaire}

\paragraph{Perceived Competence (Adapted from the Intrinsic Motivation Inventory (IMI))}
\begin{itemize}
    \item I think I performed well in making exercise recommendations during this task.
    \item This was a task that I couldn't do very well. \textit{(reverse Likert)}
    \item I believe I am skilled at suggesting suitable exercises for different individuals.
    \item After working at this task for a while, I felt pretty competent.
\end{itemize}

\paragraph{Perceived Choice --- Autonomy (Adapted from IMI)}
\begin{itemize}
    \item I felt like I had a lot of choice in deciding which exercises to recommend.
    \item I was free to choose the exercises I thought were best suited for the person described.
    \item I felt like I was strongly influenced by the AI on how to recommend exercises. \textit{(reverse Likert)}
    \item I recommended exercises in the way I wanted to.
\end{itemize}

\paragraph{Relatedness to AI (Adapted from IMI)}
\begin{itemize}
    \item I felt I could trust this AI.
    \item I felt my reasoning on this task was distant from the AI's. \textit{(reverse Likert)}
    \item I would like a chance to interact with this AI in the future.
\end{itemize}

\paragraph{Interest/Enjoyment (Adapted from IMI)}
\begin{itemize}
    \item I enjoyed this exercise recommendation task.
    \item This task did not hold my attention at all. \textit{(reverse Likert)}
    \item While I was doing this task, I was thinking about how much I enjoyed it.
    \item I thought this exercise recommendation task was a boring task. \textit{(reverse Likert)}
\end{itemize}

\paragraph{Mental Demand}
\begin{itemize}
    \item I found this task mentally demanding.
\end{itemize}

\subsection{Results: Are participants influenced by the presence of a foil?}

\begin{figure*}
     \centering
     \includegraphics[width=0.60\textwidth]{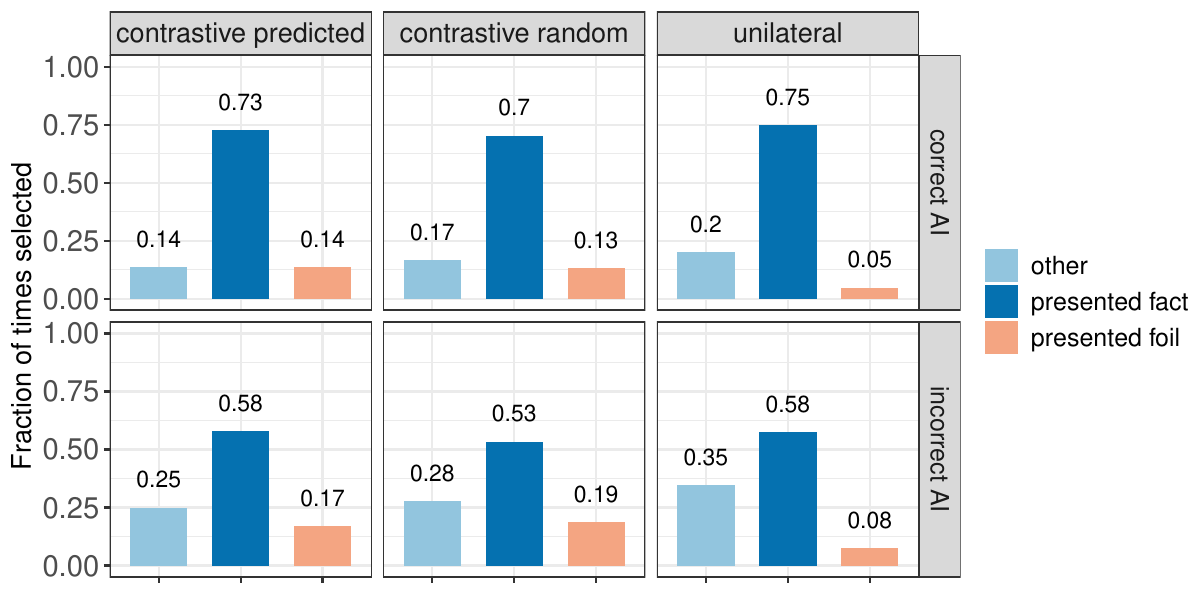}
     \caption{Are participants influenced by the presence of a foil? Comparison of answer distributions between contrastive and unilateral designs, with no foil provided in the unilateral condition.}
     \label{distribution_of_answers}
\end{figure*}

We aimed to investigate whether presenting participants with two choices influenced their likelihood of selecting the foil compared to when the foil was not visible (i.e., the unilateral condition). Our analysis revealed that when participants picked some choice other than the fact, that other choice was significantly more likely to be the foil in contrastive designs than in the unilateral condition (Figure ~\ref{distribution_of_answers}). This effect was observed in both the contrastive predicted condition, $\chi^2(2, N = 3682) = 99.30$, $p < .001$, and the contrastive random condition, $\chi^2(2, N = 3976) = 91.15$, $p < .001$. Interestingly, participants in the contrastive random and predicted conditions selected the foil at similar rates, suggesting that the mere presence of a foil, rather than its quality, influences decision-making.

\subsection{Results: Additional Analysis for Learning}

\begin{figure*}
     \centering
     \includegraphics[width=0.60\textwidth]{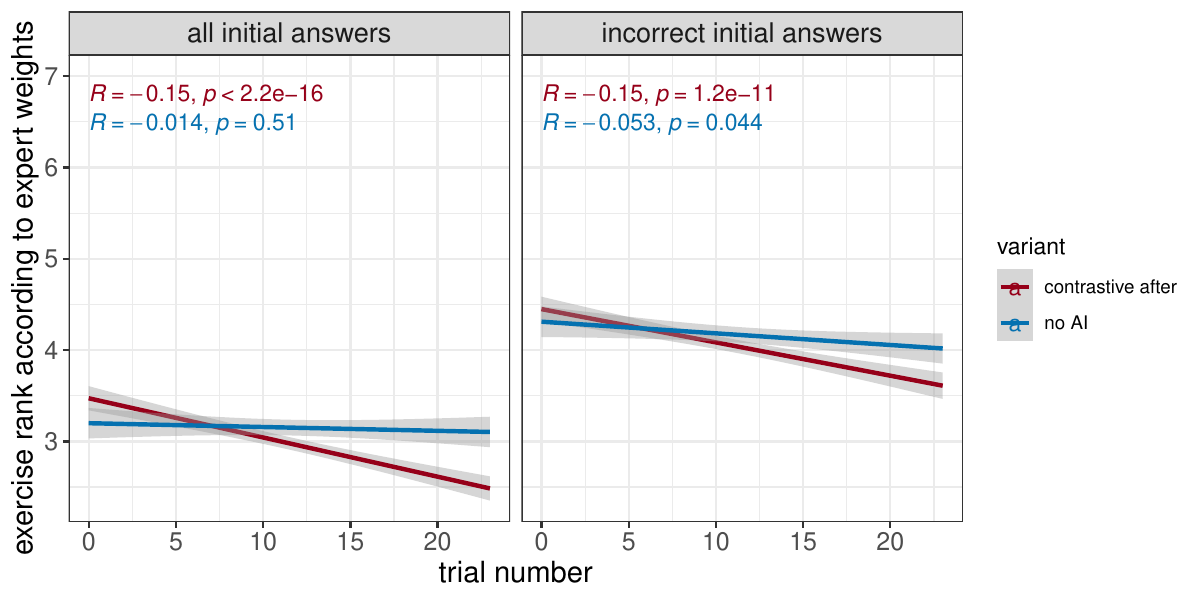}
     \caption{Correlation between trial number and rank ($1^{st}$ rank corresponding to expert's choice) for choices made without AI assistance. Participants in the contrastive after condition showed significant improvement in their initial choices over time, even when those choices were incorrect.}
     \label{contrastive_after_analysis}
\end{figure*}

To gain deeper insights into people's learning throughout the study, we conducted an additional analysis of the \emph{no AI} and \emph{contrastive after} conditions, focusing on participants' unassisted initial answers (Figure~\ref{contrastive_after_analysis}). Our results indicate that the initial choices participants provided in the contrastive after condition significantly improved over the course of the study ($r = -0.15, 95\% CI[-0.18, -0.11], p < .0001$; CI based on 1000 bootstrap samples). Even when participants provided incorrect initial guesses, those guesses progressively ranked higher according to the expert weights over time ($r = -0.15,  95\% CI [-0.19, -0.11], p < .0001$). Participants in the \emph{no AI} condition did not exhibit significant overall improvement over time ($r = -0.01$, 95\% CI [-0.05, 0.03], $p = n.s.$). However, a very weak correlation of improvement was observed for instances where they provided incorrect choices ($r = -0.05$, 95\% CI [-0.11, 0.002], $p = 0.04$).

\subsection{Results: Audit for Intervention-Generated Inequalities}
\begin{table}[ht]
\centering
\resizebox{\textwidth}{!}{%
\begin{tabular}{ll|l|l|l|l}
\hline
\textbf{Condition} & \textbf{High AOT (SE)} & \textbf{Low AOT (SE)} & \textbf{Significance} & \textbf{Effect Size (d [CI])} \\
\hline
contrastive after & 0.42 (0.03) & 0.41 (0.03) & $F_{1, 127} = 0.07, p = n.s.$ & 0.05 [-0.31, 0.40] \\
unilateral & 0.40 (0.03) & 0.38 (0.03) & $F_{1, 134} = 0.30, p = n.s.$ & 0.09 [-0.24, 0.43] \\
contrastive random & 0.46 (0.03) & 0.38 (0.03) & $F_{1, 142} = 3.77, p = 0.05$ & 0.34 [-0.01, 0.68] \\
contrastive predicted & 0.52 (0.03) & 0.42 (0.03) & $F_{1, 122} = 6.67, p = 0.01$ & 0.47 [0.11, 0.84] \\
no AI & 0.34 (0.03) & 0.31 (0.03) & $F_{1, 83} = 0.64, p = n.s.$ & 0.17 [-0.26, 0.61] \\
\hline
\end{tabular}}
\caption{ANCOVA results by condition for AOT groups, showing marginal means (SE), Significance (F-statistic, p-value), and Effect size (Cohen's d with 95\% confidence intervals).}
\label{tab:aot-results}
\end{table}

\begin{table}[ht]
\centering
\resizebox{\textwidth}{!}{%
\begin{tabular}{ll|l|l|l|l}
\hline
\textbf{Condition} & \textbf{High NFC (SE)} & \textbf{Low NFC (SE)} & \textbf{Significance} & \textbf{Effect Size (d [CI])} \\
\hline
contrastive after & 0.43 (0.03) & 0.41 (0.03) & $F_{1, 127} = 0.11, p = n.s.$ & 0.06 [-0.29, 0.41] \\
unilateral & 0.40 (0.03) & 0.38 (0.03) & $F_{1, 134} = 0.20, p = n.s.$ & 0.08 [-0.26, 0.42] \\
contrastive random & 0.42 (0.03) & 0.42 (0.03) & $F_{1, 142} = 0.02, p = n.s.$ & 0.02 [-0.30, 0.35] \\
contrastive predicted & 0.49 (0.03) & 0.47 (0.03) & $F_{1, 122} = 0.35, p = n.s.$ & 0.11 [-0.25, 0.46] \\
no AI & 0.34 (0.03) & 0.31 (0.03) & $F_{1, 83} = 0.28, p = n.s.$ & 0.11 [-0.32, 0.55] \\
\hline
\end{tabular}}
\caption{ANCOVA results by condition for NFC groups, showing marginal means (SE), Significance (F-statistic, p-value), and Effect size (Cohen's d with 95\% confidence intervals).}
\label{tab:nfc-results}
\end{table}




\end{document}